\documentclass[11pt,a4paper]{article}

\usepackage{enumitem}
 \usepackage{comment}
\usepackage[left=3cm,top=4cm,bottom=2cm,right=3cm]{geometry}
\usepackage[colorlinks=true,urlcolor = purple,linkcolor=teal,citecolor=magenta,bookmarks=true]{hyperref}
\usepackage{inputenc}
\usepackage{cite}

\usepackage{empheq} 
\usepackage{bm}
\usepackage{amsmath}
\usepackage{amsfonts,amsthm,amssymb}
\usepackage{mathtools}
\usepackage{esint}
\usepackage{graphicx}
\usepackage{xcolor}

\numberwithin{equation}{section}

\def\lt#1{\left#1}
\def\rt#1{\right#1}
\def\t#1{\widetilde{#1}}
\def\h#1{\hat{#1}}
\def\b#1{\bar{#1}}
\def\frc#1#2{\frac{#1}{#2}}

\newcommand{\p}{\partial}

\newcommand{\R}{{\mathbb{R}}}

\newcommand{\ii}{ {\rm i} }
\usepackage{authblk}

\newcommand{\ri}{{\rm i}}
\newcommand{\dd}{{\rm d}}

\newcommand{\bc}{\begin{center}}
\newcommand{\ec}{\end{center}}
\def\ba#1{\begin{array}{#1}\displaystyle}
\newcommand{\ea}{\end{array}}

\newcommand{\beq}{\begin{equation}}
\newcommand{\eeq}{\end{equation}}
\newcommand{\beqa}{\begin{eqnarray}}
\newcommand{\eeqa}{\end{eqnarray}}
\newcommand{\no}{\nonumber}
\newcommand{\n}{\nonumber\\}
\newcommand{\bi}{\begin{itemize}}
\newcommand{\ei}{\end{itemize}}
\def\lt#1{\left#1}
\def\rt#1{\right#1}
\def\t#1{\tilde{#1}}
\def\h#1{\hat{#1}}
\def\b#1{\bar{#1}}
\def\frc#1#2{\frac{#1}{#2}}

\DeclareMathOperator{\sgn}{sgn}

\newcommand{\bra}{\langle}
\newcommand{\ket}{\rangle}

\newcommand{\bs}{}
\newcommand{\ms}{\mathsf}
\newcommand{\mf}{\mathfrak}

\title{Hydrodynamic gauge fixing and higher order hydrodynamic expansion }
\usepackage{authblk}\author[1]{Jacopo De Nardis}
\author[2]{Benjamin Doyon}
\affil[1]{ Laboratoire de Physique Théorique et Modélisation, CNRS UMR 8089, CY Cergy Paris Université, 95302 Cergy-Pontoise Cedex, France.}
\affil[2]{ Department of Mathematics, King's College London, Strand WC2R 2LS, London, U.K.}
\date{}                     %% if you don't need date to appear
\begin{document}

\maketitle

\begin{abstract}
Hydrodynamics is a powerful emergent theory for the large-scale behaviours in many-body systems, quantum or classical. It is a gradient series expansion, where different orders of spatial derivatives provide an 
effective description on different length scales. We here report the first general derivation of third-order, or ``dispersive", terms in the hydrodynamic expansion. We obtain fully general Kubo-like expressions for the associated hydrodynamic coefficients, and we determine their expressions in quantum integrable models, introducing in this way purely quantum higher-order terms into generalised hydrodynamics. We emphasise the importance of hydrodynamic gauge fixing at diffusive order, where we claim that it is parity-time-reversal, and not time-reversal, invariance that is at the source of Einstein's relation, Onsager's reciprocal relations, the Kubo formula and entropy production. At higher hydrodynamic orders we introduce a more general, $n$-th order ``symmetric'' gauge, which we show implies the validity of the higher-order hydrodynamic description. 
\end{abstract}

 \tableofcontents

\section{Introduction and main results}
\label{sec:intro}

Understanding the dynamics that emerge at large scales of space in time in many-body systems is one of the most fundamental problems of theoretical physics. Hydrodynamics is arguably a powerful framework for this purpose. In recent years it has been shown to apply to a wide variety of models, chaotic or integrable, deterministic or stochastic, relativistic or not \cite{ldlandau2013,Rangamani2009,deBoer2015,Lucas2015,Lucas2018,Ku2020,PhysRevLett.122.090601,2009.06651,2006.08577,2009.13535,PhysRevLett.128.021604,Crossley2017,PhysRevD.85.085029,10.21468/SciPostPhys.12.4.130}. In its most basic description, hydrodynamics is a dynamical theory for the time evolution of local physical observables, expressed as an expansion in derivatives with respect to space and time. The convergence of this expansion has been recently discussed in relativistic field theories \cite{PhysRevLett.122.251601}, and, in quantum integrable models, the study of this expansion beyond first orders  \cite{PhysRevLett.118.226601,PhysRevX.6.041065,PhysRevLett.117.207201,SciPostPhys.2.2.014,PhysRevLett.125.240604,vir1,PhysRevB.101.180302,2005.13546,Gopalakrishnan2019,PhysRevB.102.115121,PhysRevB.96.081118,PhysRevLett.125.070601,PhysRevLett.124.210605,PhysRevLett.124.140603,Bulchandani_2019,SciPostPhys.3.6.039,Doyon_2017} is subject of current research \cite{PhysRevB.96.220302,10.21468/SciPostPhys.8.3.048,PhysRevLett.128.190401}. Hydrodynamics purports that only a restricted set of observables is necessary to describe the emergent dynamics. According to fundamental principles of statistical mechanics and ``ergodicity", these emergent dynamical degrees of freedom are associated with the densities of extensive conserved quantities admitted by the system.

In this paper, we restrict to one dimension of space for simplicity -- although there is no fundamental obstacle in generalising the theory developed here to higher dimensions. If $\mathtt q_i(x,t)$ are the values at spacetime point $x,t$ of the local conserved densities (labelled by $i$) admitted by the model, then one may expect that the one-dimensional hydrodynamic equations up to third order in the gradient expansion take the general form
\beq\label{eq:hydro-expanded}
    \p_t \mathtt q_i + \mathsf A_i^{~j}
    \p_x \mathtt q_j =  \frc12 \p_x \Big(\mathfrak D_i^{~j}
    \p_x \mathtt q_j\Big)
    \sgn(t) + \frc12 \p_x \Big(\mathcal W^{(2)}{}_i^{~j}\p_x^2 \mathtt q_j + \mathcal W^{(1,1)}{}_i^{~jk}\p_x \mathtt q_j\p_x\mathtt q_k\Big),
\eeq
where the Einstein convention of summation over repeated indices is implied. These are in fact of the form of continuity equations: indeed the flux Jacobian is of the form $\mathsf A_i^{~j} = \p \mathsf J_i/\p \mathtt q_j$, where $\mathsf J_i = \mathsf J_i(\mathtt q)$ is the Eulerian current of charge $\mathtt q_i$ (the current in the stationary, entropy-maximised state that is characterised by the values $\mathtt q_i$ of the conserved densities). 

We shall show that this higher-order hydrodynamic expansion emerges in many-body systems. All hydrodynamic coefficients, $\mathsf A_i^{~j}$, $\mathfrak D_i^{~j}$, $\mathcal W^{(2)}{}_i^{~j}$ and $\mathcal W^{(1,1)}{}_i^{~jk}$ depend on the local conserved densities $\mathtt q_k$'s. As far as we are aware, the coefficients $\mathcal W^{(2)}{}_i^{~j}$ and $\mathcal W^{(1,1)}{}_i^{~jk}$ were not known before in full generality, and we will provide explicit, Kubo-like formulae for these, written in terms of space-time integrated correlation functions of densities and currents within stationary states. Moreover, we shall show how in some cases, as in a class of interacting integrable models and in free models, it is possible to choose a set of densities such that the equation reduces to the quasilinear form
\beq\label{eq:hydro-expanded-2}
    \p_t \mathtt q_i + \mathsf A_i^{~j}
    \p_x \mathtt q_j = \frc12 \p_x \Big(\mathfrak D_i^{~j}
    \p_x \mathtt q_j\Big) + \frc12 \p_x \Big(\mathcal W^{(2)}{}_i^{~j}\p_x^2 \mathtt q_j \Big).
\eeq

For hydrodynamic equations to correctly represent the large-scale behaviour of many-body systems, just writing the naive derivative expansion \eqref{eq:hydro-expanded} is not sufficient. Two essential aspects to consider are (1) that the densities $\mathtt q_i$ of extensive conserved quantities are ambiguously defined, and this ambiguity must be lifted, and (2) that the general form of the derivative expansion is constrained by the underlying microscopic physics. In particular, the hydrodynamic coefficients must satisfy a number of relations, which are fundamental laws of many-body physics -- it is in general not sufficient to simply use, for instance, the spacetime symmetries in order to constrain the form of the hydrodynamic coefficients. These fundamental laws guarantee that the correct physical behaviours arise from the emergent hydrodynamic equation. For instance, one requires hyperbolicity of the Euler equation (the first-order expansion), the Onsager reciprocal relations for the diffusion coefficients, and non-negative entropy production at the diffusive scale -- which imply irreversibility of the flow and stability of the hydrodynamic equations. It is important to correctly fix the conserved densities in a universal, model-independent fashion, and to uncover the universal constraints on hydrodynamic coefficients that imply these fundamental laws. We shall discuss how to choose conserved densities for the hydrodynamic equation in such a way that, under a natural definition of the total entropy, not only diffusive terms lead to entropy increase, but dispersive (third order) terms do not affect it.
%We will explain how, in fact, possibly not all microscopic systems have physically meaningful third-order hydrodynamics.

%Related to this, another important question is as to evaluating, in specific models, the hydrodynamic coefficients. One often uses the specific structures of spacetime symmetries, and the transformation properties of densities and currents, to constrain the form of the hydrodynamic coefficients; for instance, in three-dimensional Galilean invariant models, up to the diffusive order, only two hydrodynamic coefficients remain to be determined from the microscopic model:  pressure and the viscosity. However, in general, such arguments are quite limited; for instance, in integrable models, generalised hydrodynamics has infinitely-many hydrodynamic coefficients, which have to be determined by other methods -- as has been done up to diffusive order \cite{2104.04462}. The problem goes beyond one of practicality: the hydrodynamic coefficients present a number of general properties, fundamental laws of many-body physics, that guarantee the correct physics of the emergent hydrodynamic equation; it is important to uncover such laws. This includes, for instance, the hyperbolicity of the Euler equation, the Onsager reciprocal relations, and non-negative entropy production at the diffusive scale, as well as the conservation of entropy for differentiable solutions at the Euler scale.

In order to frame the problem, first recall that at the Euler order, the problem is reduced to determining the exact local currents $\mathsf J_i$ in homogeneous, stationary, clustering states: these are the maximal entropy states, Gibbs or generalised Gibbs ensembles, parametrised by the ensemble averages of conserved densities $\mathtt q_k$'s. Indeed, at the Euler scale, the system is assumed to have locally maximised entropy (with respect to the available conservation laws). The function $\mathsf J_i(\mathtt q)$ is the {\em equation of state} of the thermodynamics. In fact, it is expected that the Euler scale always exists for short-range interactions, independently of the specific structure of the microscopic model \cite{1912.08496,doyon_hydrodynamic_2022}.

At the diffusive order, the local maximisation principle does not hold anymore, and the average current receives higher-derivative corrections. The Kubo formula, obtained by a linear response argument, gives $\mathfrak D_i^{~j}$ in terms of spacetime-integrated stationary state correlation functions of currents. This presents a subtlety: the total conserved quantities under Hamiltonian evolution $H$
\beq\label{conservationlaws}
    Q_i = \int_\R \dd x\,q_i(x,t), \quad \quad [H,Q_i]=0,
\eeq
give the microscopic conserved densities $q_i(x,t)$, satisfying the continuity equation
$
    \p_t q_i(x,t) + \p_x j_i(x,t) = 0,
$
only up to spatial derivatives. Thus, there is a \textit{``gauge ambiguity" at all length scales}  \cite{10.21468/SciPostPhys.6.4.049,10.21468/SciPostPhys.8.3.048}, and  the charge densities can always be redefined as 
\begin{equation}\label{eq:gaugeFixing}
    q_i(x)\mapsto q_i(x) + \p_x a_i^{(1)}(x) +  \p^2_x a_i^{(2)}(x) + \p^3_x a_i^{(3)}(x) + \ldots.
\end{equation}
This ambiguity does not affect the Euler order but gives ambiguities from the diffusive order onwards. In \cite{10.21468/SciPostPhys.6.4.049,Durnin2021} it was proposed that PT symmetry fixes, in an essential way, the first functions $a_i^{(1)}$, and that this gauge fixing extract the correct emergent degrees of freedom at the diffusive order. As we emphasise in the present paper, section \ref{sec:gaugediff}, with this gauge fixing, basic principles of statistical mechanics guarantee that these hydrodynamic coefficients admit the correct physics. In particular, the Onsager reciprocal relations hold, and entropy production is non-negative. This, we believe, is an important remark, especially in relation to some recent works investigating such aspects \cite{Casati,PhysRevE.103.052116,PhysRevE.102.030101,2208.01463}. The diffusive order of hydrodynamics is sometimes broken: the Kubo formula may give infinity, implying that superdiffusion occurs instead of diffusion. This is typical in one dimension \cite{Popkov2015,2103.01976,Jara2015,Scheie2021,Bulchandani2020,Spohn2014}, although, in the restricted case of integrable models, the diffusive order exists except for special models \cite{10.21468/SciPostPhys.9.5.075,Doyon2021}.

At the dispersive order, as far as we are aware, there is no general Kubo-like formula for the dispersive coefficients ${\mathcal W^{(2)}}_i^{~j}$, ${\mathcal W^{(1,1)}}_i^{~jk}$, and no general theory for fundamental laws they may obey. On the other hand the dispersive order of hydrodynamics is physically very relevant: for instance, we expect that dispersive hydrodynamics will describe in a universal fashion the dispersive shock waves observed in quasi-BEC \cite{Wan2006,PhysRevLett.101.170404,PhysRevA.85.033603,PhysRevA.88.013605}, free-fermion models \cite{PhysRevLett.117.130402,PhysRevB.96.220302,10.21468/SciPostPhys.8.3.048} and soliton gases \cite{https://doi.org/10.48550/arxiv.2104.05812}. In sections \ref{sec:sec3}, \ref{sec:hydroexpansion} and appendix \ref{sec:hydroexpcomp}, we obtain general, Kubo-like formulae for the dispersive hydrodynamic coefficients $${\mathcal W^{(2)}}_i^{~j}, \quad {\mathcal W^{(1,1)}}_i^{~jk}.$$ We obtain them by an appropriate nonlinear response analysis, assuming that the microscopic model is Hamiltonian (quantum or classical). We discuss in section \ref{sec:symmetricgauge} the general properties of these coefficients, in particular the choice of gauge $a_i^{(2)}$, if it exists, for the resulting conserved densities to be the correct emergent degrees of freedom and the link with a notion of entropy production that is valid up to the dispersive order. We propose a gauge fixing procedure which we denote as \textit{symmetric gauge}. In quantum integrable systems, this allows us to compute exactly the dispersive hydrodynamic coefficients, as we show (along with simple examples) in section \ref{sec:examples}. We conclude in section \ref{sec:conclu}.

\section{Diffusive hydrodynamics and PT gauge fixing}\label{sec:gaugediff}

In this section, we review some basic notions related to hydrodynamics at the diffusive scale, as well as some results and concepts first introduced in \cite{10.21468/SciPostPhys.6.4.049,Durnin2021} which we believe are important to stress. 

First, one should note that diffusion in many-body systems appears in \textit{two} main different incarnations. On the one hand, one may look at the spatial spread of two-point, two-time correlation functions in some stationary state,
\begin{equation}
   \langle  q_i(x,t)q_j(0,0)\rangle ^{\rm c} = \langle q_i(x,t)q_j(0,0)\rangle  - \langle  q_i(x,t)\rangle \langle q_j(0,0) \rangle  .
\end{equation}
Under diffusive spreading, the variance $\int_\R \dd x\,x^2( q_i(x,t)q_j(0,0))^{\rm c}$ grows linearly and we can define the coefficient of such a growth as the diffusion matrix times the covariance matrix $\mathsf C_{ij} = \int_\R \dd x\, \langle  q_i(x,t)q_j(0,0)\rangle ^{\rm c}$ 
\begin{equation}
   \mathfrak D_i^{~k}\mathsf C_{kj} = \lim_{t \to \infty} t^{-1} \int_\R \dd x\,x^2 \langle  q_i(x,t)q_j(0,0)\rangle ^{\rm c} ,
\end{equation}
where we have neglected eventual ballistic components (see below). This is a consequence of hydrodynamic diffusion, and can be obtained by linear response on the diffusive hydrodynamic equation.

On the other hand, one may look at the response of each current under an externally applied field gradient. Understanding this requires a perturbative analysis of the microscopic dynamics under small applied forces, and therefore linear response analysis of the microscopic evolution equations. Looking at the linear response of various currents $\mathtt j_i$ (and again subtracting their eventual ballistic part) under the corresponding various forces $F^i$, one gets a matrix: the Onsager matrix, 
\beq\label{onsager}
    \mathfrak L_{ij} = \frc{\p \mathtt j_i}{\p F^j}\Big|_{\boldsymbol F = 0}.
\eeq

The general theory of diffusion is mainly characterised by four (related) important relations involving these quantities: the Einstein relation, Onsager reciprocal relations, the Kubo formula, and entropy production. Expressed in the general context of multiple conservation laws, these are as follows. First, the Einstein relation is the statement that the Onsager matrix is related to the diffusion matrix, by a factor of the static covariance matrix,
\beq\label{einstein}
    \mathfrak L_{ij} = \mathfrak D_i^{~k}\mathsf C_{kj}
    \qquad
    \mbox{(Einstein relation).}
\eeq
Thus hydrodynamic diffusion, which describes large-scale dynamics, is determined by the response of currents to applied forces. Second,
the Onsager reciprocal relations state that the Onsager matrix is symmetric:
\beq\label{reciprocal}
    \mathfrak L_{ij} = \mathfrak L_{ji}\qquad
    \mbox{(Onsager reciprocal relations).}
\eeq
This is surprising and says, in words, that the response of the $i^{\rm th}$ current to the $j^{\rm th}$ force is the same as that of the $j^{\rm th}$ current to the $i^{\rm th}$ force. Third, the Kubo formula, which makes Onsager reciprocal relations explicit, relates the Onsager matrix, and thus diffusion by the Einstein relation, to time-integrated total current correlations,
\beq\label{kubo}
    \mathfrak L_{ij} = \int_{-\infty}^\infty \dd t\int_\R \dd x\,( j_i(x,t),j_j(0,0))^{\rm c}
    \qquad
    \mbox{(Kubo formula),}
\eeq
where the two-point correlation is symmetric, and in the quantum case given by the usual Kubo-Martin-Schwinger product \cite{Petz1993-yb}, see also Appendix \ref{sec:kmsproduct}.
The Kubo formula is important, showing that the Onsager matrix is non-negative. This finally implies positive entropy production under the hydrodynamic equation. That is, the total entropy of the hydrodynamic state,
\begin{equation}
   S = - {\rm Tr}[\rho_{\rm hydro} \log \rho_{\rm hydro}] , \quad \quad \rho_{\rm hydro}= e^{-\int \dd x \,\beta^i(x) q_i(x)}
\end{equation}
where $\beta^i(x)$ are the local thermodynamic potentials, $q_i(x)$ the local densities, can only increase by the formula
\beq
    \frc{\dd S}{\dd t} = \int_\R \dd x\,\p_x \beta^i
    \mathfrak L_{ij}\p_x \beta^j \geq 0
    \qquad\mbox{(entropy production)},
\eeq
which makes diffusive hydrodynamics irreversible, see sec. \ref{sec:entropyincrease}.

It is a simple matter to account for the presence of ballistic transport in addition to diffusive behaviours; this translates into a different definition of the Onsager matrix \eqref{onsager}, and the Kubo formula \eqref{kubo}, where the ballistic part is subtracted out, see in particular the next section \ref{sec:sec3}.

These important relations have been derived at various levels of generality and in various models. However, questions still arise regarding their domain of validity \cite{Casati,PhysRevE.103.052116,PhysRevE.102.030101,2208.01463}. What properties must a many-body system possess for this general theory to hold? It is often believed that time reversal invariance is required; however, Onsager reciprocal relations have been observed in systems without time-reversal invariance \cite{Casati}.

We claim that time-reversal invariance is {\em not necessary} for the Einstein relation, Onsager reciprocal relations, the Kubo formula and positive entropy production. Instead, all relations follow solely from general principles of statistical mechanics, along with the condition that {\em the system admits PT-symmetry}. The presence of PT-symmetry in a model means that there exists a choice of conserved densities $q_i(x,t)$ and an involution $\sigma$ of the algebra of observables, such that $\sigma$  flips space and time but otherwise preserves all conserved densities,
\beq\label{eq:PTsym}
    \sigma (q_i(x,t)) = q_i(-x,-t)\qquad \mbox{(PT-symmetry).}
\eeq
(Note that as a consequence of the conservation laws, PT-symmetry also preserves all currents, $\sigma (j_i(x,t))= j_i(-x,-t)$; it also preserves all maximal entropy states). That is, see explicitly eq. \eqref{eq:Onsager1} and \eqref{eq:Onsager}, 
\beqa
    \lefteqn{\mbox{PT-symmetry}\Rightarrow} &&\\ && \mbox{Einstein relation, Onsager reciprocal relations, Kubo formula, entropy production.}\no
\eeqa

We believe this important statement has not been fully appreciated in the literature.
As indeed mentioned around eq. \eqref{eq:gaugeFixing},  local conservation laws $\p_t q_i + \p_x j_i=0$ are invariant under the re-definitions, or gauge transformation, namely at diffusive order we are free to redefine the densities and their currents as 
\beq
    q_i\mapsto q_i + \p_x a_i^{(1)},\quad j_i\mapsto j_i -\p_t a_i^{(1)}.
\eeq
As shown in \cite{10.21468/SciPostPhys.6.4.049} and recalled in section \ref{sec:symmetricgauge}, PT-symmetry fully fixes the conserved densities and currents up to terms that involve two derivatives (that is, $a^{(1)}_i$ must be a total derivative with respect to space, or to any of the ``times" admitted by the model). A different choice of gauge in general not only breaks the general relations above but leads to {\em inequivalent hydrodynamic equations}. Without the PT-symmetric choice of gauge,  generically, the resulting hydrodynamic equations do not admit positive entropy growth, invalidating the use of hydrodynamics. 

Below, we assume throughout that PT symmetry holds, eq.~\eqref{eq:PTsym}. 

\section{Dispersive hydrodynamics: setup and main results}\label{sec:sec3}

In this section, we express our main results about the dispersive-order hydrodynamic equations.

We consider a (one-dimensional) Hamiltonian system, quantum or classical, with (homogenous) hamiltonian $H$ and which admits a certain number of conserved quantities with associated local conservation laws, Eq.~\eqref{conservationlaws}. The hydrodynamic variables are the values of the conserved densities $\mathtt q_i(x,t)$ at spacetime point $x,t$. To each such set of values we can associate generalised inverse temperatures $\bar{\beta}^i(x,t)$ by using averages in homogeneous, stationary maximal entropy states, which are of Gibbs form:
\beq\label{eq:hydrovariables}
    \bar{\beta}^i \leftrightarrow \mathtt q_i \quad : \quad \frac{1}{Z} {\rm Tr}(e^{-\bar{\beta}^i Q_i} q_i) = \mathtt q_i.
\eeq

When considering the hydrodynamic equation including the diffusive (and higher) orders, it is not true that the local state at $x,t$ is a maximal entropy state. Thus the $\b\beta^i(x,t)$'s, for given values of $x$ and $t$, do not describe the physical local state at $x,t$, but are just a way of parametrising the average local densities. Nevertheless, the (abstract) stationary state described by $\b\beta^i(x,t)$'s can be used not only to describe the local averages, but also the hydrodynamic coefficients as functions of $x,t$.

In such states, we thus consider one-point averages and connected two- and three-point correlation functions for local observables $a(x,t), b(y,s),\ldots$, which are denoted by
\beq
    (a) = \frc1Z {\rm Tr}(e^{-\bar{\beta}^i Q_i} a),\quad (a(x,t),b(y,s)), \quad
    (a(x,t),b(y,s),c(z,u))
\eeq
(clearly $(q_i) = \mathtt q_i$). These are symmetric under the exchange of the fields. See Appendix \ref{sec:kmsproduct} for the explicit definitions, which in the quantum case involves the Kubo-Mori-Bogoliubov (KMB) inner product and its 3-point generalisation. All hydrodynamic coefficients at $(x,t)$ are evaluated in terms of connected correlation functions evaluated in the state $e^{-\b\beta^i(x,t)Q_i}$ corresponding to the local values of the $\mathtt q_i(x,t)$'s (thus are functions of the $\mathtt q_i(x,t)$'s).

We start by defining the standard static covariance matrices (or susceptibility matrices) 
\beq\label{CBmatrix}
    \mathsf C_{ij} = (Q_i,q_j) = -\frc{\p(q_i)}{\p\b\beta^j},\quad
    \mathsf B_{ij} = (J_i,q_j)=
    -\frc{\p(j_i)}{\p\b\beta^j},
\eeq
the 3rd order covariance is
\beq\label{C3matrix}
    \mathsf C_{ijk}' = (Q_i,Q_j,q_k)
    = \frc{\p^2(q_i)}{\p\b\beta^j\p\b\beta^k},
\eeq
and the flux Jacobian is
\beq\label{Amatrix}
    \mathsf A_i^{~j}
    = \mathsf B_{ik}\mathsf C^{kj}
    = \frc{\p(j_i)}{\p(q_j)}.
\eeq
Here and below, we use the upper-index shorthand for the inverse $\mathsf C$ matrix
\beq
    \mathsf C_{ij}\mathsf C^{jk}
    = \delta_i^k,
\eeq
as well as Einstein's notation of implied summation over repeated upper and lower indices.

We will show in section \ref{sec:hydroexpansion} and appendix \ref{sec:hydroexpcomp} that, by evaluating the form of the currents as functions of the densities after local relaxation using microscopic nonlinear response theory, the hydrodynamic expansion of the current  leads to the following expression for the currents  $\mathtt j_i(x,t)$ of the hydrodynamic variables $\mathtt q_i(x,t)$ at point $x$
\begin{eqnarray}
	\mathtt j_i &=& ( j_i) - \frc12 {\mf L}_{ij}^{(1)} \ms C^{jk} \p    \mathtt q_k\,
    \sgn(t)
	+\, \frc12\mf L_{ij}^{(2)}\p^2 \b \beta^j
	+\,\frc12  \mf L_{ijk}^{(1,1)} \p\b\beta^j
	\p\b\beta^k. \label{eq:hyjfinal}
\end{eqnarray}
Note how the diffusive part of the current has a factor $\sgn(t)$. This characterises the time direction in which evolution is taken: expression \eqref{eq:hyjfinal} is to be put within the conservation law giving the hydrodynamic equation, which is assumed to evolve from its initial condition at time 0. The hydrodynamic current takes different forms for a microscopic evolution towards positive times ($\sgn(t)=1$) and towards negative times ($\sgn(t)=-1$), because in both cases, entropy should increase. This is a sign of irreversibility.

The diffusive Onsager matrix derived from the hydrodynamic expansion reads in full generality as 
\begin{equation} \label{eq:Onsager1}
  {\mf L}^{(1)}_{ij} =\lim_{t\to \infty}2\int \dd y \ y [ (j_i(y,t) ,q_j) - \ms A_i^{~k} (q_k(y,t) ,q_j)  ]  .
\end{equation}
Here and below, when no arguments appear, the observable is evaluated at $(0,0)$. Expression \eqref{eq:Onsager1}, in general, is neither symmetric nor positive definite, and does not satisfy Einstein's relation. By imposing PT symmetry, it can be recast into the standard Kubo-like formula 
\begin{equation}\label{eq:Onsager}
  {\mf L}^{(1)}_{ij} =\int_{-\infty}^{\infty} \dd s \ (J_i(s),j_j)^C ,
\end{equation}
where $J_i(s) = \int \dd y\,j_i(y,s)$ and the superscript ${}^C$ indicates that the infinite time limit of the correlator is subtracted out:
\beq
    (J_i(s),j_j)^C =
    (J_i(s),j_j) - \lim_{t\to\infty}
    (J_i(t),j_j)
    = (J_i(s),j_j) - 
    \mathsf A_i^{~k}(Q_k,j_j),
\eeq
where hydrodynamic projection has been used in the last equality. Now expression \eqref{eq:Onsager} is symmetric, and can be shown to be positive semi-definite.

At the dispersive order, we have the two following higher-order coefficients 
\beqa\label{eq:maincoeff2}
    \mf L^{(2)}_{ik}
    &=& \Big[-\int \dd x\,x^2 (j_i(x,t) - \mathsf A_i^{~j}q_j(x,t),q_k(0,0))\Big]_{t^0}\n
    \mf L^{(1,1)}_{ilm} &=&
    \Big[\int \int \dd x\dd y\,xy (j_i(0,t) - \mathsf A_i^{~j}q_j(0,t),q_l(x,0)q_m(y,0))\Big]_{t^0}.
\eeqa
The symbol $\big[\cdot \big]_{t^0}$ means that one should only retain the coefficient of order  $O(t^0)=O(1)$ in an asymptotic expansion in integer powers of $t$ at large $t$. Assuming that the asymptotic expansion has a finite number of positive powers, this can be written explicitly as limits with subtractions of the positive powers by solving the recursive system
\beq
    \big [f(t)\big]_{t^n}
    = \lim_{|t|\to\infty}
    f_n(t)/t^n,\quad
    f_n(t)
    = f(t) - t^{n+1} \big[f(s)\big]_{s^{n+1}},
\eeq
starting with $f_{n+1}(t)=f(t)$ if $f(t) = O(t^n)$. In fact, in general, the asymptotic expansion may be different as $t\to\infty$ and $t\to-\infty$; we account for this by keeping factors of $\sgn(t)$ in $\big [f(t)\big]_{t^n}$.  As we will see, the expressions within the square brackets above have integer-power asymptotic expansion; for $\mf L^{(2)}_{ik}$, the expression is $O(t)$, and for $\mf L^{(1,1)}_{ilm}$, it is $O(t^2)$, and we may also write 
\begin{equation}
  {\mf L}^{(1)}_{ij}\sgn(t) =\Big[2\int \dd y \ y [ (j_i(y,t) ,q_j) - \ms A_i^{~k} (q_k(y,t) ,q_j)  ]\Big]_{t^0}  .
\end{equation}

Eq.~\eqref{eq:hydro-expanded} is obtained from the above (higher-order) Onsager matrices by setting:
\beq\label{eq:maincoeff1}
   \mathfrak{D}_i^{~j} =  {\mf L}^{(1)}_{ik} \mathsf C^{kj}  ,\quad {\mathcal W^{(2)}}_i^{~j}
    =  \mf L^{(2)}_{ik}\mathsf C^{kj},\quad {\mathcal W^{(1,1)}}_i^{~jk}
    = - (\mf L^{(1,1)}_{ilm} + {\mathcal W^{(2)}}_i^{~n}\mathsf C'_{nlm})\mathsf C^{lj}\mathsf C^{mk}.
\eeq
Equations \eqref{eq:hydro-expanded}, \eqref{eq:maincoeff2} and \eqref{eq:maincoeff1} are our main results. Further, two related results are as follows.

First, the equation \eqref{eq:hydro-expanded} also implies equations for the evolution of the two or higher point functions. For two-point functions in  space-time stationary states, we apply hydrodynamic linear response theory by taking the functional derivative $\delta/\delta\beta^i(0,0)$ on the hydrodynamic average, and the homogeneous limit. We find for the Fourier transform $S_{qq}(k,t)= \int \dd x\,e^{\ri kx}( q(x,t),q(0,0))$:
\beq\label{eq:2pointExpansion}
    S_{qq}(k,t) =
    \exp\Big[\mathsf A\ri k t + \frc12\mathfrak D (\ri k)^2|t| +
    \frc12\mathcal W^{(2)} (\ri k)^3t\Big]
    \ms C  .
\eeq
For higher-point functions, one should generalise the Euler-scale hydrodynamic projection theory to higher derivative orders, and also consider potential long-range correlations \cite{2206.14167}; we hope to come back to this in future work.

Second, we shall then show in section \ref{sec:symmetricgauge} that the imposition of a \textit{2nd order symmetric gauge} guarantees positive entropy increase for eq.~\eqref{eq:hydro-expanded} and a simpler definition of the hydrodynamic coefficients 
\beqa\label{eq:maincoeff3}
    \mf L^{(2)}_{ik}
    &=&  \mf L^{(2)}_{0,ik} = \lim_{t \to 0}\Big[-\int \dd x\,x^2 (j_i(x,t) - \mathsf A_i^{~j}q_j(x,t),q_k(0,0))\Big] \n
    \mf L^{(1,1)}_{ilm}
    &=& \mf L^{(1,1)}_{0,ilm} = \lim_{t \to 0}
    \Big[\int \int \dd x\dd y\,xy (j_i(0,t) - \mathsf A_i^{~j}q_j(0,t),q_l(x,0)q_m(y,0))\Big].
\eeqa
Note how the extraction of the $O(t^0)$ coefficient in a large-time expansion has been replaced by the simple limit $t\to0$ (the limit can be taken from above or from below, giving the same result by PT symmetry). The 2nd order symmetric gauge may or may not exist; this will depend on the details of the model under study. We will show that it exists for the free fermion model, and provide indications that it does also for the hard-rod model.

\section{Hydrodynamic expansion}\label{sec:hydroexpansion}

We now explain the main steps in the derivation of the dispersive terms in the results Equations \eqref{eq:hyjfinal}, \eqref{eq:maincoeff2} and \eqref{eq:maincoeff1}, from which the hydrodynamic expansion \eqref{eq:hydro-expanded} follows by the continuity equations. We do this using microscopic response theory, as introduced in \cite{Durnin2021}, which allows us to correctly describe the local relaxation necessary for the emergence of the hydrodynamic equation. All technical calculations are reported in appendix \ref{sec:hydroexpcomp}.

The technique relies on the ansatz that the relation between the average currents $\mathtt j_i$ and densities $\mathtt q_i$, within the $n^{\rm th}$-order hydrodynamic approximation, can be obtained as the relation that arises for such quantities in the large-time limit, from a density matrix $\rho_{\rm hydro}$ of the form
\beq\label{eq:initial}
	\rho_{\rm hydro} = \exp\Big[-
	\int \dd x\,\sum_i \beta^i(x) q_i(x)
	\Big]/Z,
\eeq
expanded to $n^{\rm th}$ order in derivatives of the potentials $\beta^i(x)$. When trying to evaluate the large-time limit of the currents and densities {\em after} doing the expansion in derivatives of potentials, order by order, one obtains divergences; this is because the true large-time limit is non-perturbative in potential derivatives. Thus the large-time asymptotic expansion of these quantities has growing terms. However, we will show that the expression of large-time currents in terms of large-time densities and their spatial derivatives, {\em has a finite and well-defined large-time limit}: all growing terms cancel. This fact was already known up to the 2nd order, and here we show it at the 3rd order of derivatives in the potentials. The resulting expression of currents as functions of densities and their spatial derivatives is therefore valid at the timescale {\em when ``partial" relaxation has occurred} (mesoscopic timescale). Indeed, it is natural to define the timescale when partial relaxation has occurred as the timescale controlled by the vanishing terms in the large-time asymptotic expansion of currents and densities, as these are expected to be of microscopic origin. The resulting expression for the currents is then the correct one to use in the hydrodynamic expansion, as microscopic scales are much smaller than the timescale associated with the hydrodynamic evolution; this expression expresses the local relaxation that occurs within ``fluid cells".

Note that the finiteness of the currents when written in terms of densities and their spatial derivatives at large times, at the third order of the derivative expansion, is a nontrivial statement; this is one of the main technical achievements of this paper.

For the argument, it is sufficient to consider $\rho_{\rm hydro}$ to be at time $t=0$, and to evaluate currents and densities at the point $x=0$.

Expanding the density matrix, we write
\beq\label{eq:hydrostateexpansion}
	\rho_{\rm ini} = \exp\Big[
	-\int \dd x\,\sum_i (\beta^i + x\p\beta^i + \frc{x^2}2 \p^2\beta^i) q_i(x)
	\Big]/Z.
\eeq
Here and below, for convenience, we denote $ f =  f(0)$, $\p f = \p_{x} f(x)|_{x=0}$, $\p^2 f = \p_x^2 f(x)|_{x=0}$. Notice that the potentials $\beta^i(x)$ are \textit{not the same} as the $\bar{\beta}^i(x)$ in \eqref{eq:hydrovariables}; the $\bar{\beta}^i(x)$ are formal  potentials associated values of conserved densities $\mathtt q_i$ (here, obtained in the large-time limit). In contrast, the $\beta^i$ are the initial values of the potentials used to study the relaxation process. The two sets of potentials are related to each other via a derivative expansion, $\beta^i = \b\beta^i + O(\p\b\beta)$.

The relaxation process is studied by expanding \eqref{eq:hydrostateexpansion}, evaluating the currents and densities order by order, and taking the large time limit of their (microscopic) time evolution.

In the following, we shall lighten the notation by discarding indices.  We use $(\cdot)$ for the expectation value in the state $Z^{-1}e^{- \beta^i Q_i}$, and likewise $(,)$ and $(,,)$ are the appropriate connected, KMB-type correlators (see Appendix \ref{sec:kmsproduct}) in that state. With this, we define the integrated correlators
\beq\label{eq:basicU}
	\bs U^{(n)} = -\int \dd y\,y^n ( \bs q(0,t),\bs q(y,0)),\quad
	\bs U^{(n,m)} = \int \dd y\,y^n z^m( \bs q(0,t),\bs q(y,0),\bs q(z,0))
\eeq
and
\beq\label{eq:basicV}
	\bs V^{(n)} =  -\int \dd y\,y^n ( \bs j(0,t),\bs q(y,0)),\quad
	\bs V^{(n,m)} = \int \dd y\,y^n z^m( \bs j(0,t),\bs q(y,0),\bs q(z,0)).
\eeq
We are interested in $\b o(0,t) = Z^{-1}{\rm Tr}\big( \rho_{\rm hydro} o(0,t)\big)$ with the state \eqref{eq:hydrostateexpansion}. The expansion takes the form
\beqa \label{eq:expansiontotmain} \label{initj}
	\b{\bs j}(0,t) &=& ( \bs j) + \bs V^{(1)}\p\bs\beta + \frc12 \bs V^{(2)} \p^2\bs\beta + \frc12 \bs V^{(1,1)} (\p\bs\beta,\p\bs\beta), \n
	\b{\bs q}(0,t) &=& ( \bs q) + \bs U^{(1)}\p\bs\beta + \frc12 \bs U^{(2)} \p^2\bs\beta + \frc12 \bs U^{(1,1)} (\p\bs\beta,\p\bs\beta),\n
	\overline{\p \bs q}(0,t) &=&  \bs U^{(0)}\p\bs\beta + \bs U^{(1)} \p^2\bs\beta + \bs U^{(1,0)} (\p\bs\beta,\p\bs\beta),\n
	\overline{\p^2 \bs q}(0,t) &=&  \bs U^{(0)}\p^2\bs\beta + \bs U^{(0,0)} (\p\bs\beta,\p\bs\beta).
\eeqa
Here, we use the notation $(\p \beta, \p \beta)$ to denote contraction respect to different indices: $ (,{}, ) (\p \beta , \p \beta) \equiv \sum_{m,n} ( ,{}_n,{}_m)\p \beta^n \p \beta^m$, to be contrasted with $(,{}) \p^2 \beta\equiv  \sum_{m}(,{}_m) \p^2 \beta^m  $, etc. We also used the symmetry $\bs U^{(1,0)}_{ijk}=\bs U^{(0,1)}_{ikj}$. The remaining non-contracted index is the same on the right- and left-hand sides.

The relaxation process is described by expressing the values of the currents in the first line of \eqref{eq:expansiontotmain} as functions of the hydrodynamic variables  \eqref{eq:hydrovariables} and their spatial derivatives, i.e.
\beq\label{basicini}
	\mathtt q_i \equiv \b q_i(0,t),\quad
	\p\mathtt q_i \equiv \overline{\p q}_i(0,t),\quad
	\p^2\mathtt q_i \equiv \overline{\p^2 q}_i(0,t).
\eeq
As explained, this step is fundamental in order to obtain finite results at the mesoscopic timescale. We simply expand the integrated correlators as asymptotic series in large $t$, taking care of re-expressing all integrated correlators as functions of $\b\beta^i$'s instead of $\beta^i$'s, the former defined via \eqref{eq:hydrovariables}.  Again, quite remarkably, {\em under the assumption that asymptotic expansions of integrated correlators at large times take integer powers}, in the large (mesoscopic) time limit the currents $\mathtt j_i = \b j_i(0,t)$ become finite expressions of $\mathtt q_i,\p\mathtt q_i,\p^2\mathtt q_i$ (all positive-power, time-divergent terms cancel). The computations are reported in app.~\ref{sec:hydroexpcomp}, giving our main results. It would be interesting to extend these ideas to situations where large-time asymptotic expansions take non-integer powers (giving superdiffusion and related phenomena).

\section{Gauge fixing}\label{sec:symmetricgauge}

As mentioned in the introduction, a full specification of the 3rd-order hydrodynamics requires an appropriate gauge fixing, namely a precise choice of the hydrodynamic densities $q_i(x)$, see eq. \eqref{eq:gaugeFixing}. In this section, we introduce the symmetric gauge. This is done for a quantum theory, but we expect a similar concept to exist also for classical theories. We propose that the symmetric gauge, if it exists, is the correct gauge. We justify this statement by showing that hydrodynamics constructed under the symmetric gauge has a \textit{positive increase of the space-integrated local entropy of the hydrodynamic state} (more precisely, the 3rd-order term in the hydrodynamic equation does not affect the entropy, thus the entropy increase from the diffusive term, if it is present, still is dominant and positive).

Recall that as we are looking at higher orders in the hydrodynamic derivative expansion, the assumption that the system has locally maximised entropy, and therefore is locally described by a Gibbs state, is not valid anymore. Hence, the notion of local entropy cannot simply be taken as the von Neumann entropy of the local Gibbs state.
We here introduce an appropriate and natural definition of local entropy; the result that the total entropy may only increase (implying irreversibility of the flow) justifies why this definition is useful.

We emphasise that, for any given system, {\em there is no a priori guarantee that the symmetric gauge, or any gauge that implies irreversibility (and therefore stability) of the hydrodynamic flow, should exist}. That is, there may not be local conserved densities that satisfy the conditions of the symmetric gauge. If the gauge doesn't exist, one may argue that the 3rd-order hydrodynamic is simply not a valid description of the large-scale physics, and additional degrees of freedom, beyond the conserved densities $q_i(x)$, must be introduced to characterise the state at a given time. We shall see that in some cases there must exist a space of local densities under the symmetric gauge.

\subsection{The symmetric gauge}

Let us consider the eigenstates $\{|e \rangle \}_e$ of the (homogenous) Hamiltonian $H$,  which also diagonalises the total momentum operator $K$, with momentum  and energy eigenvalues $k_e, \varepsilon_{e}$. We consider the matrix elements of the charge densities at a given position $x=0$, written as 
$\langle e | q_i(0) |e' \rangle$. By translation invariance, 
\begin{equation}
    \langle e | q_i(x) |e' \rangle= e^{\ri x (k_{e'} - k_{e})} \langle e | q_i(0) |e' \rangle,
\end{equation}
and, as we consider the system of infinite size and on continuous space, the total momentum $k_e$ takes all values in $\R$. 
Notice that the eigenstate parametrisation $\{ e \}$ may be fully characterised by different quantities beyond energy $\varepsilon_{e}$ and momentum $k_e$, as the latter can, in principle, be highly degenerate.
Under the choice of such a parametrisation, we then define charges $q_i$ in the \textit{n-th order symmetric gauge} such that their matrix elements do not depend on the momentum difference up to corrections of order $(k_{e'}-k_{e})^{n+1}$, i.e.
\begin{equation}\label{eq:conditionsymmetric}
    \frac{ \p^{\ell}}{\p ( {k_{e'}-k_e)^\ell}}   \ \langle e | q_i(0) | e' \rangle \Big|_{k_{e'}-k_{e}=0}=0 \quad \forall\; \ell \leq n.
\end{equation}
Here $\p / \p(k_{e'}-k_e) = \frc12(\p/\p k_{e'} + \p/\p k_e)$, and all other parameters in the choice of parametrisation of the state $|e\ket$ are kept fixed under differentiation.

This gauge may always be formally achieved by using the $n$-th order gauge shifts (see eq. \eqref{eq:gaugeFixing})
\begin{equation}\label{eq:gf2}
   q_i(x)\mapsto \t q_i(x) = q_i(x) + \sum_{\ell=1}^{n} \p^{\ell}_x a_i^{(\ell)}(x) .
\end{equation}
Indeed, we may choose the $a_i^{(\ell)}(x)$ to cancel the momentum-difference dependence of the matrix elements up to corrections of order $O[(k_e- k_{e'})^{n+1}]$,
\begin{equation}\label{eq:syGd}
   \langle e | \t q_i(0) |e' \rangle = \langle e |  q_i(0) |e' \rangle + \sum_{\ell=1}^n \Big[ \ri(k_{e'}- k_{e}) \Big]^\ell  \langle e | a^{(\ell)}_i(0) |e' \rangle ,
\end{equation}
by choosing appropriately their matrix elements for $k_e\approx k_{e'}$. In fact, this can be expressed quite explicitly using the center-of-mass operator $X$ conjugate to the total momentum $[X,K]=\ri$, which acts as
\begin{equation}\label{Xder}
    X | e\rangle =  -\ri \frac{\p }{\p k_e} | e \rangle.
\end{equation}
The operators $a^{(\ell)}_i$ can be constructed recursively as multiple anticommutators with the centre of mass operator $X$, by cancelling the leading power of $k_{e'}-k_e$ at each step. We define $q_i^{(\ell)} = q_i^{(\ell-1)} + \partial_x  a_i^{(\ell)}$ with $q_i^{(0)} = q_i$ and $q_i^{(n)} = \t q_i$, and we have
\begin{equation}\label{eq:gaugefixingfunctions}
  a^{(\ell)}_i(0)=  -\frac{1}{ 2^\ell \ell !} \underbrace{\{ X, \{ \ldots,  \{ X,}_{\ell}    q_i^{(\ell-1)}(0) \}\} \}.
\end{equation}
In fact, by the recursive procedure, one can see that this equality is required, at each step, only for matrix elements {\em at equal momenta $k_e = k_{e'}$}.
{

For example, imposing the second-order symmetric gauge reads as 
\begin{equation}
   \langle e | \t q_i(0) |e' \rangle = \langle e |  q_i(0) |e' \rangle +     \ri(k_{e'}- k_{e})    \langle e | a^{(1)}_i(0) |e' \rangle  -      (k_{e'}- k_{e})^2    \langle e | a^{(2)}_i(0) |e' \rangle ,
\end{equation}
where
\begin{equation}
   \langle e | a^{(1)}_i |e' \rangle =  \ri  \frac{\partial}{\partial ( k_{e'} - k_{e})}   \langle e | q_i |e' \rangle,
\end{equation}
and
\begin{align}
  &  \langle e | a^{(2)}_i |e' \rangle =   \frac{\partial^2}{\partial( k_{e'} - k_{e})^2} \langle e | q_i + \partial_x a_i^{(1)} |e' \rangle \nonumber \\& = \frac{\partial^2}{\partial( k_{e'} - k_{e})^2} \langle e | q_i | e' \rangle - \frac{\partial^2}{\partial( k_{e'} - k_{e})^2} ( k_{e'} - k_{e}) \frac{\partial}{ \partial ( k_{e'} - k_{e})}\langle e |  q_i |e' \rangle.
\end{align}
We refer to as \textit{fully symmetric gauge} the case where all functions $a_i^{(\ell)}$ are fixed for all $\ell>0$, and matrix elements of charge densities do not depend at all on the total momentum difference, namely the $n=\infty$ order in eq. \eqref{eq:syGd}. \\

Two caveats are in order about this formal construction. First, the operator $X$ is not uniquely determined by $[X,K] = \ri$. Only relation \eqref{Xder} fixes it. Thus the choice of $X$ is related to the choice of parametrisation $\{ e\}$ of the eigenstates $e$ for the extra quantum numbers not fixed by the total momentum and energy. Second, in general, there is no guarantee that the multiple anticommutator expression \eqref{eq:gaugefixingfunctions} gives a local operator $\tilde{q}_i$. Yet, the locality of the densities is (presumably) important for the hydrodynamic equation to make sense. For instance, in a system of $N$ particles with positions $\h x_i$ and momenta $\h p_i$, the total momentum operator is $K = \sum_i \h p_i$, and a natural choice is $X = N^{-1}\sum_i \h x_i$; but the anticommutators of $X$ with local observables are not generically local. Perhaps one may find an appropriate choice of $X$ (thus of eigenstates parametrisation) for the result to be local. More likely, the condition \eqref{eq:gaugefixingfunctions} is only satisfied for $k_e\approx k_{e'}$, and the matrix elements beyond this neighbourhood are fixed by requiring locality. In order to argue that it should be possible to keep locality after gauge fixing, recall that locality is assessed by considering the commutators (here, we concentrate on diagonal matrix elements for simplicity of the discussion) 
\begin{equation}
  \bra e |[q_i(x),q_j(0)]|e\ket = 2\, {\rm Im}\sum_{e'}   e^{\ri (k_e-k_{e'})x} \langle e| q_i | e' \rangle \langle e' | q_j | e \rangle 
\end{equation}
and requiring the vanishing of this commutator at large enough distances. Thus, if the charges $q_i$ after being gauged remain local, all such commutators should decay at large $x$. The formally-written sum over $e'$ includes an integral over the continuous values of $k_{e'}$, and a typical mechanism for the vanishing of commutators is to perform a contour deformation of this integral, towards imaginary directions, up to the positions of singularities of the matrix elements or the integration measure. The imaginary part of the closest such position determines the exponent in the (expected) exponential decay. This mechanism suggests that it is the analytic structure of the matrix element of $\langle e | q_i | e' \rangle$ away from $k_e=k_{e'}$ that determines the locality of $q_i$, hence generically, we should expect that it is possible to modify form factors in a neighbourhood of $k_e=k_{e'}$ without affecting the locality of the densities $q_i(x)$.

We remark that the PT gauge choice we discussed earlier is a {\em weaker condition} than the symmetric gauge, but the modification by the gauge functions $a^{(\ell)}_i$, as constructed above, \textit{does not} affect the PT symmetry of the charge densities. Thus, all conclusions that follow from PT symmetry are unchanged. Indeed, as shown in \cite[App C2]{10.21468/SciPostPhys.6.4.049} 
PT gauge fixing is unique up to a shift $\partial_x z(x)$, where $z(x)$ is a PT antisymmetric operator. 
As $\sigma(X) = -X$, any shift functions $a_i^{(\ell)}$ constructed by eq. \eqref{eq:gaugefixingfunctions} 
has PT-parity equal to $(-1)^\ell$, i.e.  $\sigma(a_i^{(\ell)}(x)) = (-1)^\ell  a_i^{(\ell)}(-x)$,  hence it does not affect PT-symmetry of $q_i$. Notice moreover that for every $\ell$ odd, all diagonal matrix elements must vanish, $\bra e|a^{(\ell)}_i|e\ket=0$, and thus $a^{(\ell)}_i$ must be a total, either space of time, derivative.

For a consistent hydrodynamic theory at diffusive (2nd) order, PT symmetry is sufficient, as we have argued above (and see \cite{10.21468/SciPostPhys.6.4.049}). At dispersive (3rd) order, as we now argue, the 2nd order symmetric gauge is enough. In fact, it implies that integrated correlators appearing in the hydrodynamic expansion \eqref{eq:expansiontotmain} are all \textit{polynomial functions in $t$} (with in general different polynomials for the expansion as $t\to\pm\infty$ as per  time-reversal symmetry / anti-symmetry), and in particular   
\begin{equation}\label{eq:zeroentropycond}
    \mf L^{(2)} =\mf L^{(2)}_{0}, \quad  \mf L^{(1,1)} = \mf L^{(1,1)}_0,
\end{equation}
where the coefficients on the right-hand side are obtained by computing the integrated correlators directly at time $t=0^+$. This, as we show in sec.~\ref{sec:entropyincrease},  guarantees a positive entropy increase.

Let us argue for eq. \eqref{eq:zeroentropycond}: 
First, let us notice the important fact that matrix elements of densities and currents indeed satisfy the continuity equation
\begin{equation}
    (\varepsilon_{e} - \varepsilon_{e'}) \langle e | q_i(0) | e' \rangle =  (k_e - k_{e'}) \langle  e | j_i(0) | e' \rangle ,
\end{equation}
therefore, modulo a zero density of states, the \textit{matrix elements of the charge densities $q_i$ between states with zero momentum difference $(k_e - k_{e'})=0$ are zero unless their energy difference is also zero $ (\varepsilon_e - \varepsilon_{e'})=0$}. We then insert a resolution of identity $\sum_{e'} |e' \rangle \langle e' |$ in the computation of one of the correlation functions appearing in  \eqref{initj} 
\begin{equation}
    [V^{(2)}]_{ik} = \sum_e \sum_{e'}\rho_{e,e'} \delta''(k_e-k_{e'}) e^{- i t (\varepsilon_e - \varepsilon_{e'}) } \langle e | q_i |e' \rangle \langle  e | j_k |e' \rangle^*.
\end{equation}
The measure $\rho_{e,e'}$ is the KMS density of states obtained after integration over the imaginary time. 
In the 2nd order symmetric gauge, the derivative of momenta can not act on the matrix elements, as they do not depend on $k_e-k_{e'}$, therefore it can only act on the measure $\rho_{e,e'}$ or the energy phase $e^{- i t (\varepsilon_e - \varepsilon_{e'}) }$, but keeping the constraint that then the sum over $e'$ only involves states with zero energy difference $\varepsilon_e - \varepsilon_{e'}=0$. We conclude that the correlator must be a \textit{polynomial} in $t$. One can then easily repeat the same argument for the integrated correlators $ U^{(2)}$, as well as $V^{(1,1)}$ and $U^{(1,1)}$. \\

In quantum integrable models, it is known how to define matrix elements between stationary states, defined by quasiparticle densities, and their excitations, defined by particle-hole shifts \cite{CortsCubero2019,De_Nardis_2015,10.21468/SciPostPhys.6.4.049,10.21468/SciPostPhys.9.6.082}. There, every charge density $q_i(0)$ acts on a quasiparticle state specified by its density $| \rho_{\theta }  \rangle$ of quasiparticles, in the following manner 
\begin{equation}
    q_i (0)| \rho_{\theta } \rangle = \sum_{e'}  F^{(q_i)}_{e'} | \rho_{\theta } , e' \rangle,
\end{equation}
where $ F^{(q_i)}_\mu $ are usually called finite-density form factors and the excitations $e'$ are typically in the form of particle-hole, ({$\rm ph$}), namely a set of $h_i$, the holes, are replaced by a new set $p_i$, the particles, in the finite-density thermodynamic state $| \rho_{\theta } \rangle$ (characterised by dressed momentum and energy $k(\theta)$, $\varepsilon(\theta)$ and particle filling $n(\theta)$) and the summation over states become a sum over particle-hole contributions
\begin{equation}
 \sum_{e'} = \sum_{m=1}^{\infty} \frac{1}{(m!)^2}\Big[\prod_{i=1}^m  \int \frac{dp_i}{2\pi} k'(p_i) (1-n(p_i)) \int  \frac{dh_i}{2 \pi} k'(h_i) n(h_i)\Big] .
\end{equation}
The only non-zero measure set of  states with zero momentum and energy with finite matrix elements are in the one and two-particle hole sectors. Indeed we have, for states differing by one particle-hole excitation, energy and momentum difference reads
\begin{equation}
{\rm 1ph} \quad \quad  \varepsilon(p) - \varepsilon(h) \quad, \quad    k(p) - k(h),
\end{equation}
and zero momentum implies directly zero energy. 
For two particle holes instead, energy and momentum difference reads
\begin{equation}
  {\rm 2ph} \quad \quad    \varepsilon(p_1) + \varepsilon(p_2) - \varepsilon(h_1) - \varepsilon(h_2) \quad, 
    \quad    k(p_1) + k(p_2) - k(h_1) - k(h_2).
\end{equation}
Here clearly zero momentum difference does not directly imply zero energy difference, but there exist a non-zero measure set of states where energy also is zero, namely those states where the difference $p_1-h_1$ and $p_2-h_2$ (or equivalently $p_1-h_2$ and $p_2-h_1$) are sent to zero. These states are known to have diverging terms in their matrix elements that compensate their zeros at small momentum (giving this way to finite matrix elements with zero momentum and energy). Their contribution contains the diffusive parts of the integrated correlation function \cite{10.21468/SciPostPhys.6.4.049,RevCorrelations,Cubero2019}. At higher particle-hole numbers there is no such divergence at zero energy and momentum (modulo a zero density of states) and therefore form factors with zero momentum are strictly zero. Moreover, by imposing a $n$-th symmetric gauge, they are zero up to $O( (k_n- k_m)^{n+1})$ corrections, and therefore they can be neglected in the calculation of any correlation function of the type as in \eqref{initj}
\begin{equation}
\int \dd y  \ y^n (q_i(y,t),o),
\end{equation}
(with a generic operator $o$) and their higher points generalisation. This statement leads to the computation of the coefficient \eqref{eq:w2interacting}, which is reported in sec. \ref{sec:FFfreeexp2}.

Finally a comment on the existence of the symmetric gauge at order larger than $1$ (the first order is simply PT symmetry as specified above). It is clear that in any model where the center of mass operator $X$ can be written we can gauge its densities as in \eqref{eq:gaugefixingfunctions}. This is doable in any model on the continuum, where momentum can always be defined. For models on the lattice or generically with broken translational symmetry we cannot prove the existence of such a gauge, even if it could be effectively possible to introduce a quasi-conserved momentum operator or a non-local one (as is the case in interacting quantum integrable models).

\subsection{Entropy production }\label{sec:entropyincrease}
The increase, over hydrodynamic times, of the total entropy, provides a measure of the validity of the hydrodynamic equation. Indeed, the total entropy, evaluated as the sum over all hydrodynamic fluid cells of each cell's entropy, can only increase, as the information that the state contains, as seen as a set of separate fluid cells, can only decrease. This is irreversibility of the hydrodynamic flow. Intuitively, a negative entropy increase would signify that information from microscopic (neglected) degrees of freedom is coming up to the hydrodynamic state, invalidating its use.  This is why choosing the appropriate gauge for the densities $q_i(x)$ (that is, the appropriate total-derivative terms, see \eqref{eq:gaugeFixing}) is fundamental to describing the state properly at the hydrodynamic scale. Only those sets of densities (not necessarily unique) that guarantee positive entropy increase are valid for a correct hydrodynamic description.

This however brings the question of the correct definition of the total entropy of fluid cells. At the Euler scale, this is straightforward, as each fluid cell is assumed to have maximised entropy, with $\b\beta^i(x,t)$ describing the local Lagrange parameters. The von Neumann entropy of each maximal-entropy state is the natural candidate, and the correct one, giving zero entropy increase at the Euler scale. At the diffusive scale, the states in fluid cells are not maximal entropy states, as expectations of local observables admit first-derivative corrections. Thus, (as explained in previous sections) the $\b\beta^i(x,t)$'s do not describe actual maximal entropy states, but are just a way of parametrizing the local conserved densities $\mathtt q_i(x,t)$ (which themselves describe the local state and solve the hydrodynamic equation). Nevertheless, it turns out that the same (naive) definition of the local entropy -- the von Neumann entropy associated to $\b\beta^i(x,t)$'s -- also correctly gives entropy increase.

At the third order, however, this would be too naive. In this subsection, we explain what the correct definition is, which in principle should be valid at all orders; why it reproduces the naive one at diffusive order; and why, in the symmetric gauge, the third-order (dispersive) contribution to the entropy increase vanishes, thus guaranteeing (with entropy increase from the diffusive order) the correct behaviour.

Consider the time-slice at time $t$. The state is described by the local density averages $\mathtt q_i(x,t):x\in\R$. By the map \eqref{eq:hydrovariables}, it is equivalently described by $\b\beta^i(x,t):x\in\R$. But also, the state may be described by the following density matrix on time slice $t$:
\beq\label{eq:hs}
	\rho_{\rm hydro}(t) = Z^{-1}\exp\Big[-
	\int \dd x\, \beta^i(x,t) q_i(x,t)
	\Big],\quad
	\mathtt q_i(x,t) = \bra q_i(x,t)\ket_{\rho_{\rm hydro}}\,.
\eeq
The expansion in the 2nd equation of \eqref{initj}, {\em at $t=0$} (and with $\b q(0,t)$ replaced by $\mathtt q(0,0)$ on the left-hand side), provides the relation between $\mathtt q_i(0,0)$, and $(q_i)$ (the GGE average of $q_i$ in the state described by $\beta^j(0,0)$'s) and the derivatives $\p\beta^j(0,0),\,\p^2\beta^j(0,0)$. This expansion can be inverted, and, translating to arbitrary points $x,t$, in this way one determines the $\beta^j(x,t)$'s in \eqref{eq:hs} in terms of $\mathtt q_i(x,t)$'s and their spatial derivatives. The state \eqref{eq:hs} is thus an effective description of the time-slice $t$ up to the appropriate derivative order (here: the dispersive order). Our proposal is simply that the von Neumann entropy of this state is in fact the correct total hydrodynamic entropy,
\begin{equation}
    S_{\rm hydro} (t) = -{\rm Tr}[\rho_{\rm hydro}(t)\log \rho_{\rm hydro}(t)].
\end{equation}
We show that his can only increase under hydrodynamic time evolution, up to, including, the dispersive order. We first note that a direct calculation gives
\begin{align}
   & \dot S_{\rm hydro}(t) =  \int \dd x\,\beta^i(x,t)\left(\dot {\mathtt q_i}(x,t) - \bra \dot q_i(x,t)\ket_{\rho_{\rm hydro}} \right) \nonumber \\&= \int \dd x\,\p\beta^i(x,t)\,(\mathtt j_i(x,t) - \bra j_i(x,t)\ket_{\rho_{\rm hydro}}).
\end{align}
The hydrodynamic current $\mathtt j_i(x,t)$ is given by \eqref{eq:hyjfinal} (at $t>0$ as we are looking for entropy increase towards positive times). The local ``current deformation" $\bra j_i(x,t)\ket_{\rho}$ is obtained, at $(0,0)$, from the 1st equation in \eqref{initj} again evaluated at $t=0$ (and with $\b j_i(0,t)$ replaced by $\bra j_i(0,0)\ket_{\rho}$ on the left-hand side); and at other points $(x,t)$ simply by translation. Every quantity should be re-written in terms of $\mathtt q_i(x,t)$, equivalently $\b\beta^i(x,t)$.

Let us denote the quantities defined in \eqref{eq:basicU}, \eqref{eq:basicV} at $t=0$ using the subscript $\bullet_0$. By PT symmetry (see also \eqref{eq:U1is0}) one immediately obtains
\beq\label{u1v10}
    [U^{(1)}_0]_{ij} =0,\quad [V^{(1)}_0]_{ij} = 0.
\eeq
The first equation simplifies the transformation from $\beta^i$ to $\b\beta^i$, obtained using the 2nd equation in \eqref{initj}:
\beq
    \b\beta^i = \beta^i - \frc12 \mathsf C^{il}[U^{(2)}_0]_{lj}\p^2 \b\beta^j
    -\frc12 \mathsf C^{il}[U_0^{(1,1)}]_{ljk}\p\b\beta^j\p\b\beta^k.
\eeq
The second equation in \eqref{u1v10} simplifies the expression for the current deformation,
\begin{equation}
    \langle {\bs j_i} \rangle_{\rho_{\rm hydro}} = ( \bs j_i) + \frc12 [{V}^{(2)}_0]_{ij} \p^2\b\beta^j + \frc12 [{V}^{(1,1)}_0]_{ijk} \p\b\beta^j\p\b\beta^k.
\end{equation}
Putting everything together, we then finally obtain the following total entropy increase:
\beq\label{eq:entropyincreaseFinal}
 \dot{S} = \dot{S}_{\rm diff} + \frac{1}{2}  \int \dd x\,  \p\b\beta^i \Big[
	  ( \mf L^{(2)} -\mf L_0^{(2)})_{ij} \p^2\b\beta^j 
	+
	\Big(  \mf L^{(1,1)} -  \mf L^{(1,1)}_0\Big)_{ijk} \p\b\beta^j \p\b\beta^k
	\Big] ,
\eeq
where, with subscript $\bullet_0$, we have the 3rd order coefficients computed directly at $t=0$, see eq. \eqref{eq:maincoeff3}. The entropy increase generated by diffusive terms is given as usual by
\begin{equation}\label{eq:diff}
    \dot{S}_{\rm diff} = \frc12 \int \dd x \ \p\b\beta^i [\mathfrak{L}^{(1)}- \mathfrak{L}_0^{(1)}]_{i,j}  \p\b\beta^j = \frc12 \int \dd x \ \p\b\beta^i [\mathfrak{L}^{(1)}]_{i,j}  \p\b\beta^j  .
\end{equation}

We note that at the dispersive order, we do not expect integrated correlators to show discontinuities at $t=0$: the limits $t\to0^\pm$ agree and thus the definitions of $\mf L_0^{(2)}$ and $\mf L^{(1,1)}_0$ in \eqref{eq:maincoeff3} make sense. Therefore, in the symmetric gauge, where the polynomial form implies that the $O(t^0)$ terms are given by the evaluation at $t=0^+$, see \eqref{eq:zeroentropycond}, we indeed have the equalities \eqref{eq:maincoeff3}, and therefore the vanishing of the dispersive entropy-increase term in \eqref{eq:entropyincreaseFinal}. In the absence of symmetric gauged densities, the integral in Eq.~\eqref{eq:entropyincreaseFinal} is nonzero; as this integral does not have a specific sign, in this case the entropy increase due to dispersive terms (3rd order in spatial derivatives) may take both positive and negative values. Making the dispersive contribution zero is the only way which, in combination with positive semi-definiteness of the Onsager matrix $\mathfrak{L}^{(1)}$ (which follows from PT symmetry), guarantees positive entropy production, hence the stability of the dispersive hydrodynamic equation. This is achieved in particular if the set of local densities $q_i(x)$ is chosen such that the integrated correlators determining the currents at 2nd order in derivatives, are polynomial functions in $t$, implying that  condition \eqref{eq:zeroentropycond} holds. This is indeed true under the $2$nd order symmetric gauge.

We stress that the continuity of correlators $\mathfrak{L}^{(2)}$ and $\mathfrak{L}^{(1,1)}$ at $t=0$, is not shared by the diffusive order correlators: even if the integrated correlators $[V^{(1)}]_{ij}$ are polynomial in $t$, the term of order $O(t^0)$ is antisymmetric under time reversal, hence is proportional to $\sgn(t)$ and the limit $t\to0^\pm$ disagree. Also, the time-0 integrated correlator, $[V^{(1)}_0]_{ij}$, is exactly 0, see eq.~\eqref{u1v10}, by PT symmetry. Hence, the correlator $ [V^{(1)}]_{ij}$ (and the same for $[U^{(1)}]_{ij}$) is \textit{discontinuous} 
\begin{equation}
   \lim_{t \to 0^+} [V^{(1)}]_{ij} \neq  [V_0^{(1)}]_{ij}=0.
\end{equation}
Therefore, given that  $\mathfrak{L}^{(1)} = \lim_{t \to 0^+} [V^{(1)} - A U^{(1)}]$, we have  $[\mathfrak{L}^{(1)} - \mathfrak{L}_0^{(1)}] = \mathfrak{L}^{(1)} \neq 0 $   and diffusion still leads to finite, yet positive, entropy increase, also under symmetric gauge.

\section{Examples of dispersive hydrodynamics in integrable models}\label{sec:examples}

In this section, we provide some examples of applications of our hydrodynamic theory, in particular for certain integrable systems.

\subsection{Free fermions gas}\label{sec:FreeF}

We here consider the hydrodynamic description of a fermionic model (with fermionic number conservation) on the continuum space given by the generic Hamiltonian 
\begin{equation}
    H =  \int \frac{\dd k}{2 \pi} \varepsilon(k) c^\dagger_k c_k,
\end{equation}
with dispersion velocity $v(k) = \partial_k \varepsilon(k)$. 
The densities of charges can be defined using the well-known Wigner function description \cite{PhysRev.40.749,Moyal1949,PhysRevD.20.414}, namely
\begin{equation}\label{eq:chargesFF}
    q_i(0)= \int \dd y  \int \frac{\dd k}{2 \pi} \mathsf h_i(k) e^{i y k} c^\dagger_{-y/2} c_{y/2} =  \int \frac{\dd k_1}{2 \pi} \frac{\dd k_2}{2 \pi}   \mathsf h_i((k_1 + k_2)/2) c^\dagger_{k_1} c_{k_2},
\end{equation}
with $\ms h_i(k)$ the single-particle eigenvalue of the charge $Q_i=\int \dd x \ q_i(x)$ (this definition can be directly extended to systems on a lattice, provided $\int \dd y $ is replaced by $\sum_{y \in \mathbb{Z}}$ \cite{Essler2022}). 
  The definition of the charges \eqref{eq:chargesFF}, with the shift in $y$ symmetric between the two fermionic operators, respect the fully symmetric gauge, as the matrix elements are not functions of $k_1-k_2$ at all orders in $k_1-k_2$.

It is known that under this gauge, in free fermionic theories, can allow for a hydrodynamic expansion at all orders, as shown in \cite{PhysRevB.96.220302,10.21468/SciPostPhys.8.3.048}. However we can limit ourselves to a $2-$nd order symmetric gauge (higher orders won't affect the hydrodynamic equation up to dispersive order) and in section \ref{sec:FFfreeexp}, we shall show how hydrodynamic expansion gives the following expression for the 3rd order hydrodynamic coefficients, on a generic state specified by the occupation function $n(k)$
\begin{equation}\label{eq:zzfree}
  \mathfrak{L}^{(2)}_{ik} =  \frac{1}{12}\int \frac{\dd k}{2 \pi} v''(k) n(k)(1-n(k)) \ms h_i(k) \ms  h_l(k)  , \quad \quad    \mf L^{(1,1)}
    = -\mathcal W^{(2)} \mathsf C',   
\end{equation}
with 
\begin{equation}\label{eq:freefermionsResult}
    \mathsf C _{il}= \int \frac{\dd k}{2 \pi}   n(k)(1-n(k)) \ms h_i(k) \ms h_l(k) ,    \quad  \quad   \mathcal W^{(2)}_{i}{}^l  = \frac{1}{12}\int \frac{\dd k}{2 \pi} v''(k) \ms  h_i(k) \ms h_l(k).
\end{equation}
Therefore, the hydrodynamic expansion leads to the same  well-known hydrodynamic equation for the expectation value $n(k,x,t)$ of the occupation (Wigner) function \begin{equation}
    \hat{n}(k,x,t) =  \int \dd y  e^{i y k} c^\dagger_{x+y/2} c_{x-y/2} ,
\end{equation}
that reads as 
\begin{equation}
    \partial_t n(k,x,t) = - v(k) \partial_x n(k,x,t) + \frac{v''(k)}{24} \partial_x^3 n(k,x,t),
\end{equation}
in full accord with previous derivations \cite{PhysRevLett.110.060602,Viti2016,Essler2022}. 

\subsection{Gauge ambiguity and free fermions vs interacting systems}

Free fermions systems satisfy the condition \eqref{eq:zeroentropycond}, namely zero entropy increasing, for all choices of gauges, as their charge densities are quadratic in fermionic operators, namely all their charge densities are characterised by zero entropy increase (this is true at diffusive scale, where $\mathfrak{L}^{(1)}=0$ for free theories, no matter what gauge is chosen to fix the functions $a^{(1)}_i(x)$, \cite{Spohn2018}). The choice of the fully symmetric gauge, therefore, is merely convenient, and it does not affect the validity of the hydrodynamic description, see sec. \ref{sec:freeFGauge} for additional details, however \textit{it does affect the form of the hydrodynamic equations}. In particular, a different choice of densities leads to a non-symmetric $\mathfrak{L}^{(2)}$ matrix and the violation of relation \eqref{eq:3pointdisconnect}. We denote by $q_i$ the densities in the (2nd-order) symmetric gauge and by $\t q_i$ a set of (still PT invariant) charges not in the symmetric gauge. The two are related by a shift $\partial_x^2 a^{(2)}$. When describing the hydrodynamic state with the hydrodynamic variables $ \mathtt q_i = {\rm Tr} [q e^{- \bar{\beta^i} Q_i }]/Z $, our result of \eqref{eq:hyjfinal} for symmetric gauged densities gives
\beq\label{eq:first}
    \p_t \mathtt q_i=   \mathsf A \mathsf C
    \p \bar{\beta}   - \frac{1}{2}  \p_x (\mf L^{(2)}   \p^2 \bar{\beta} )  , 
\eeq 
while using the charges $\t q = q - \partial_x^2 a^{(2)}$ we have a different hydrodynamic equation indeed
\beq \label{eq:second}
    \p_t \tilde{\mathtt q}_i=   \mathsf A \mathsf C
    \p \tilde{\bar{\beta}}  - \frc12 \p_x \Big(  \, \frc12 \tilde{ \mf L}^{(2)}\p^2  \tilde{\bar{\beta}}
	+\,\frc12 \tilde{ \mf L}^{(1,1)} (\p \tilde{\bar{\beta}} ,
	\p\tilde{\bar{\beta}}) \Big),
\eeq
The fact that different choices of charge densities lead to different hydrodynamic equations here is resolved by noticing that  eq. \eqref{eq:second} is merely the evolution of the expectation value $\langle \tilde{q} \rangle = \langle q - \partial_x^2 a^{(2)} \rangle = \mathtt q_i   - \partial_x^2 \langle  a^{(2)} \rangle $ in the hydrodynamic states evolving under \eqref{eq:first}. Indeed the evolution of this latter is given by 
\begin{equation}
  \partial_t  \tilde{\mathtt q}_i   =  \mathsf A \mathsf C
    \p \bar{\beta}  + \mathsf A \partial_x (\partial_x^2 \langle a^{(2)} \rangle ) -   \frac{1}{2}  \p_x (\mf L^{(2)}   \p^2 \bar{\beta} )  =   \mathsf A \mathsf C
    \p \bar{\beta}  -  \frc12 \p_x \Big(  \, \frc12 \tilde{\mf L}^{(2)}\p^2  \bar{\beta}
	+\,\frc12  \tilde{\mf L}^{(1,1)} (\p\b\beta,
	\p\b\beta) \Big),
\end{equation}
where we used that in the second term on the right-hand side of eq. \eqref{eq:second} we can change $\tilde{\bar{\beta}} \to {\bar{\beta}}$ at this hydrodynamic order, and 
where the extra terms are generated by the term $\partial_x^2 \langle a^{(2)} \rangle$, and they read 
\begin{equation}
    \tilde{\mf L}^{(2)} = \mf L^{(2)} - \ms A \frac{\delta^2  \langle a \rangle}{\delta \bar{\beta}^2 } 
\end{equation}
\begin{equation}
    \tilde{\mf L}^{(1,1)} =   - \ms A \frac{\delta    \langle a \rangle}{\delta \bar{\beta} \delta \bar{\beta} } .
\end{equation}
Therefore the two different hydrodynamics are consistent, and there is no privileged choice for the gauge functions $a_i^{(2)}$ in free fermionic systems.

The same argument would instead fail in a generic interacting model due to the finite diffusive terms, i.e. with finite Onsager matrix $\mathfrak{L}^{(1)}$ in the expression of the current \eqref{eq:hyjfinal}. In generic interacting models indeed gauge must be properly fixed, and the hydrodynamic equations for densities in different gauges would be inconsistent with each other, as only the proper choice of charge densities, i.e. the set that guarantees positive entropy increase, can be used to construct stable hydrodynamic theories.

 \subsection{Interacting quantum integrable system}

In {quantum integrable models}, where in past years the theory of generalised hydrodynamics have been extensively developed for Euler and diffusive hydrodynamic, see for example \cite{PhysRevX.6.041065,PhysRevLett.117.207201,SciPostPhys.2.2.014,PhysRevB.104.115423,PhysRevB.106.134314,PhysRevLett.123.130602,PhysRevLett.125.240604,vir1,PhysRevB.101.180302,2005.13546,Gopalakrishnan2019,PhysRevB.102.115121,PhysRevB.96.081118,PhysRevLett.125.070601,PhysRevLett.124.210605,PhysRevLett.124.140603,Bulchandani_2019,SciPostPhys.3.6.039,Doyon_2017}, the 2nd order symmetric gauge also guarantees a relation between the three-point coefficients and the two-point ones, see app.~\ref{sec:FFfreeexp2}, 
\begin{equation}\label{eq:3pointdisconnect}
       \mf L^{(1,1)}
    = -\mathcal W^{(2)} \mathsf C',
\end{equation}
which implies the hydrodynamic equation for charges in such gauge, 
\beq
    \p_t \mathtt q_i + \mathsf A_i^{~j}
    \p_x \mathtt q_j = \frc12 \p_x \Big(\mathfrak D_i^{~j}
    \p_x \mathtt q_j\Big) + \frc12 \p_x \Big(\mathcal W^{(2)}{}_i^{~j}\p_x^2 \mathtt q_j \Big)  ,
\eeq
with hydrodynamic coefficients that can be analytically computed. In such models, indeed hydrodynamic variables can be related to a density of quasiparticles $\rho(\theta)$
\begin{equation}
    \mathtt q_i(x,t) = \int \dd \theta\, \rho(\theta;x , t) \ms h_i(\theta) \quad \forall i ,
\end{equation}
where $\ms h_i(\theta)$ is the one-particle eigenvalue of the total charge $Q_i$. The relation involves the momentum $k(\theta)$, energy $\varepsilon(\theta)$, filling $n(\theta)= 2\pi \rho(\theta)/k'(\theta)$, scattering shift $T(\theta-\theta')$ and effective velocity $v^{\rm eff}(\theta)= \varepsilon'(\theta)/k'(\theta)$, as it is  the case for interacting and free integrable models. The coefficients, still under second-order symmetric gauge, are then expressed as operators in the space of functions of $\theta\in\mathbb{R}$, and the dispersive terms turn out to be a simple generalisation of the free fermionic ones of eq. \eqref{eq:freefermionsResult}, reading as 
\begin{equation}\label{eq:w2interacting}
 \mathcal  W^{(2)} =   (1-  nT )^{-1} \Big[ \frac{k' (\theta)\varepsilon'''(\theta) - k'''(\theta) \varepsilon'(\theta)}{12(k'(\theta))^4}\Big] (1 - n T),
\end{equation}
analogously to the other (known) coefficients
\begin{equation}
    \mathsf A =  (1-  nT )^{-1}   v^{\rm eff}  (1 - n T) , \quad \mathsf C =  (1-  nT )^{-1}   \rho  (1-n)  (1 -  T n)^{-1},
\end{equation}
giving the 3rd order generalised hydrodynamic equation for the density $\rho(\theta;x,t)$ is
\begin{equation}
    \partial_t \rho + \partial_x (v^{\rm eff} \rho) = \frac{1}{2} \partial_x ( \mathfrak{D} \partial_x \rho) + \frac{1}{2} \partial_x (\mathcal  W^{(2)} \partial^2_x \rho),
\end{equation}
and with the diffusion kernel $\mathfrak{D}$ already been derived in previous works and reported for example in \cite{DeNardis2019}.

\subsection{Dispersive hydrodynamics of the hard rods gas}

A gas made of hard rods particles with fixed length $a$ represents the simplest implementation of an integrable model, with a constant microscopic scattering shift defined as $T=-a/2\pi$, and a well-defined hydrodynamic limit \cite{Spohn1982-yq}, well discussed in these past years  
\cite{2004.07113,CaoCradle}. 
In particular, as shown in \cite{Spohn1982-yq}, in such a classical system on a ring of size $L$, with total density of rods $\bar{\rho}$, is possible to express the two-point correlator of the density of particles $n(v,x,t)=L^{-1} \sum_i \delta(x- x_i) \delta(v-v_i)$, with velocity $v,v'$, at all orders in $k$ on a stationary state 
\begin{equation}
    S_{v,v'}(k,t) = \int \dd x e^{i k x} ( n(v,x,t) , n(v',0,t) ) \quad\quad  S(k,t) = e^{t \mathsf W(k)} \mathsf C(k),
\end{equation}
which can then be expanded at low momentum, analogously to our result of eq. \eqref{eq:2pointExpansion}, as 
\beq
	S(k,t)
	=\exp\Big[\ri k\mathsf A t - \frc{k^2}2 \mathfrak D |t| - \frac{1}{2}\ri \mathcal W^{(2)} k^3 t \Big]\mathsf C,
\eeq
with the convective matrix 
\begin{equation}
   [\mathsf A](v,v')= \frc{1}{1- a \bar{\rho}} \delta(v-v') v + \frac{\bar{\rho} a}{(1- a \bar{\rho})^2} v h(v) -  \frac{\bar{\rho} a}{(1- a \bar{\rho})} v' h(v),
\end{equation}
 the diffusion kernel
\begin{equation}
    [\mathfrak D](v,v')=   \frc{\bar{\rho} a^2}{1- a \bar{\rho}} ( \delta(v-v')  r(v)  - |v-v'| h(v) ),
\end{equation}
and the dispersive one
\begin{equation}\label{eq:WHR}
  [ \mathcal W]^{(2)}(v,v') = \frac{1}{12} \Big(  \frac{a^3 \bar{\rho}}{1- a \bar{\rho}} v \delta(v-v') -  \frac{a^3 \bar{\rho}}{1- a \bar{\rho}} h(v) v' - \frac{a^3 \bar{\rho} (2 - a \bar{\rho})}{1 - a \bar{\rho}} h(v)v \Big).
\end{equation}
These coefficients cannot be obtained directly from our eq. \eqref{eq:w2interacting}. In this model indeed dressed energy and momentum are simple linear functions of the velocity, and their third-order derivatives with respect to $v$ are therefore zero. While instead, the convective and diffusive coefficient present a form which is entirely generic, valid for both quantum and classical integrable models, (see in \cite{10.21468/SciPostPhys.6.4.049} how they can be written for quantum models and  the hard rod gas) our result for the dispersive coefficient of eq. \eqref{eq:w2interacting} is only valid for quantum models. One possible way to extend our results of higher order hydrodynamic terms of quantum theories to the classical hard rod system would be to take a classical limit of a quantum theory (for example, as done in \cite{Koch2022}), such as the Lieb-Liniger model \cite{PhysRev.130.1605}. In this model, with interaction $c$, the scattering shift is indeed given by  
\begin{equation}
    T(\theta,\alpha) = \frac{1}{\pi} \frac{c}{(\theta-\alpha)^2 + c^2} 
\end{equation}
At large $c$ it gives $T \to \frac{1}{\pi c}$ and therefore it corresponds to the one of the Hard rod gas with $c= - 2/a$. We can therefore compute the derivative of velocity and momentum in the first non-trivial order. We have, in an equilibrium state with a density of quasiparticles $\rho(\theta)$,
\begin{equation}
    \varepsilon' = 2 \theta + T \star (n \varepsilon'), \quad \quad   k' =1 + 2 \pi T \star  \rho,
\end{equation}
where $T$ acts as a convolution kernel in the space of functions, as usual. 
We have then
\begin{equation}
    \varepsilon''' =     T'' \star (n \varepsilon'), \quad \quad  k''' =  2 \pi T'' \star  \rho.
\end{equation}
Expanding in the first non-trivial order in $1/c  = - a/2$ and imposing a zero momentum state we find 
\begin{equation}
    \varepsilon''' \mapsto
    \frac{a^3}{4 \pi} \int \dd \theta n(\theta) \varepsilon(\theta)'=0 \quad \quad  k''' \mapsto   \frac{a^3}{2  } \bar{\rho},
\end{equation}
with the bare energy mapping to one of the Hard rod gas 
   $ \varepsilon' = 2 \theta  \mapsto v$.
We can then apply the result of \eqref{eq:w2interacting}, 
giving this way a result equivalent to \eqref{eq:WHR}. We leave for future works a comprehensive analysis of the validity and the specificity of such limit from the Lieb-Liniger model to the hard rod gas.

\section{Conclusions }
\label{sec:conclu}

We have reported an extensive and complete derivation of third-order terms of the hydrodynamic gradient expansion. The new hydrodynamic coefficients are expressed as generalised Kubo formula, in terms of spatially integrated correlation functions in stationary states. We have shown that a proper gauge choice is necessary to fix the hydrodynamic charge densities, the symmetric gauge at $2$nd order, in order to guarantee positive entropy production and therefore the validity of the hydrodynamic theory. We have shown how under the proper gauge the newly introduced hydrodynamic terms can be computed in the case of quantum integrable models, extending this way the theory of generalised hydrodynamics to include dispersive terms.

Many questions are now in sight: first, we expect that numerous studies shall be devoted to computing the newly derived dispersive terms in quantum integrable systems, where some work will be also necessary to obtain the charge densities in the second-order symmetric gauge. Such dispersive terms are expected to be relevant in regimes of low temperatures and large gradients, and for example could provide a microscopic derivation of the quantum pressure in the Bose gas and of quantum hydrodynamic terms beyond linear Luttinger liquids \cite{1509.08332}, i.e. beyond Euler spreading of quantum correlations \cite{PhysRevLett.124.140603}. Then further studies shall focus on the quantum-classical correspondence of the hydrodynamic terms. As we have shown, the quantum and the classical third-order coefficients differ, at least when looking at the classical gas of hard rods, as opposed to the ballistic and diffusive terms, which instead take a universal form for both quantum and classical theories (where, at diffusive order, quantum / classical-ness is encoded only within the free energy function). In the near future, we aim to analyze more specifically the origin of such discrepancy and the type of semi-classical limits necessary to derive dispersive hydrodynamic terms for classical integrable theories from their quantum ones. 

Finally, we have also emphasise how PT symmetry is important at the diffusive order, but not sufficient at the dispersive one. In this context, it is worth mentioning that one of the most important lines of current research in emergent large-scale behaviours is the extension of hydrodynamic theories to active systems. There, the breaking of PT symmetry, manifested by non-reciprocity, plays a crucial role \cite{You-reciprocal,Fruchart-reciprocal}, and the general ideas we have developed may help to establish guiding principles for active-systems hydrodynamics.

\section*{Acknowledgment}
We acknowledge discussions with H. Spohn, F. Essler, B. Bertini, M. Fagotti, and J. Durning, who contributed at the early stages of the project.  Some of his work was performed at Aspen Center for Physics, which is supported by National Science Foundation grant PHY-1607611 and at Galileo Galilei Institute during the scientific program ``Randomness, Integrability, and Universality''. The work of JDN has been partially funded by the ERC Starting Grant 101042293 (HEPIQ). The work of BD was supported by the Engineering and Physical Sciences Research Council (EPSRC) under grants EP/W000458/1 and EP/W010194/1.

\appendix

\section{Kubo-Mori-Bogoliubov inner product and its generalisations} \label{sec:kmsproduct}

We are interested in correlation functions
\beq
    \bra a_1a_2\ldots\ket =
    \frc{{\rm Tr} (e^{-\beta^iQ_i} a_1a_2s)}{
    {\rm Tr} e^{-\beta^iQ_i}},
\eeq
and in particular in connected correlation functions $\bra a,b,\ldots\ket^{\rm c}$; note that $\bra a,b,\ldots\ket^{\rm c} = \bra a - \bra a\ket,b - \bra b\ket, \ldots\ket$ for the 2- and 3-point connected correlation functions.

In classical models, correlation functions (where ${\rm Tr}$ is replaced by the ensemble average) have the property of being invariant under permutations of the observables. In quantum models, this symmetry is broken, because observables are operators, and don't necessarily commute with each other.

The quantities that arise from perturbation theory in classical models are indeed such correlation functions. However, in the quantum case this is not so. Instead, correlations where operators are integrated over imaginary time are involved. In particular, with $\hbar=1$, it is the standard ``KMB inner product" that arises, which takes the form
\begin{equation}
    (a,b)_{\rm KMB} = \int_0^1 d\tau \,\langle a^{\tau} b\rangle,
\end{equation}
where $a^\tau = e^{\tau\beta^iQ_i}a e^{-\tau\beta^iQ_i}$. At the third order, it is a 3rd order ``KMB correlation function", which takes the form
\beq \label{eq:fullsymcorr} 
    ( a,b,c)_{\rm KMB} =  \frac{1}{2} \int_0^1\dd \tau\int_0^\tau\dd \tau'\,\Big[\bra a^\tau b^{\tau'}c\ket 
    +
    \bra b^\tau a^{\tau'} c\ket\Big].
\eeq
Again, we are interested in the connected versions $(a,b)_{\rm KMB}^{\rm c}$ and $(a,b,c)_{\rm KMB}^{\rm c}$, defined as above but in terms of connected correlation functions. In order to account for both classical and quantum models simultaneously, we will simply define
\beq
    (a,\ldots) = \lt\{\ba{ll}
    \bra a,\ldots \ket^{\rm c} & \mbox{(classical)}\\
    (a,\ldots)_{\rm KMB}^{\rm c} & \mbox{(quantum)}
    \ea\rt.
\eeq

It turns out that these expressions are now fully symmetric, including in the quantum case:
\beq\label{fullsymm}
    (a,b) = (b,a), \quad ( a,b,c) = ( b,c,a),\quad
   ( a,b,c) = ( b,a,c),
\eeq
(the latter two equalities are two actions of the symmetric group $S_3$ which generate the whole group, so are sufficient to show that the expression is fully symmetric). Symmetry of the KMB two-point function is well known, and the last equality in \eqref{fullsymm} is immediate. Hence, let us prove is the second equality in \eqref{fullsymm}. We only consider the first term in \eqref{eq:fullsymcorr}, as this will imply a similar result for the second term:
\beqa
\int_0^1\dd\tau\int_0^\tau \dd\tau'\,
\bra a^\tau b^{\tau'}c\ket &\stackrel{\rm KMS}=&
\int_0^1\dd\tau\int_0^\tau \dd\tau'\,
\bra b^{\tau'}c a^{\tau-1}\ket\n &\stackrel{\rm stationarity}=&
\int_0^1\dd\tau\int_{0}^{\tau} \dd\tau'\,
\bra b^\tau c^{\tau'}a\ket,
\eeqa
where in the first equality we employed the Kubo–Martin–Schwinger (KMS) relation \cite{bratteli_operator_1987}.

In order to see how these definitions apply to perturbation theory, let us consider an expansion of the stationary state under small perturbations. We may use the general relation
\begin{equation}
    e^{A + \lambda B} = e^A + e^{  A} \int_0^1 d\tau \,e^{-\tau A} \lambda B e^{ \tau A}   + e^{  A}\int_0^1 d\tau \,e^{-\tau A} \lambda B \int_{0}^{\tau} d\tau'\, e^{(\tau - \tau') A}  \lambda B e^{ \tau' A}
    + O(\lambda^3).
\end{equation}
Let us apply this to expand the following (un-normalized) density matrix in powers of the derivatives $\beta^{k\prime}$ and $\beta^{k\prime\prime}$, which is relevant to the hydrodynamic expansion:
\begin{equation}
  \rho =   \exp\left(-\int \dd x\left(\beta^k+\beta^{k\prime}x+\beta^{k\prime\prime}\frac{x^2}{2}\right)q_k(x)\right).
\end{equation}
Denoting $\rho_{\rm GGE} = e^{-\beta^kQ_k}$, we obtain
\begin{align}
    \rho  &= \rho_{\rm GGE} \,\Big( 1 - \int_0^1 d \tau  \int \dd x\, \beta^{k\prime}   x q_k^\tau(x)    +  \int_0^1 d\tau \int_{0}^{\tau} d\tau' \int \dd x \dd y \,xy   \beta^{k\prime}\beta^{l\prime}  q_k^\tau(x)  q_l^{\tau'}(y) \nonumber  \\&
    +\, \frac{1}{2}  \int_0^1 d \tau  \int \dd x \,x^2 \beta^{k\prime\prime} q_k^\tau(x)
    \Big).
\end{align}
Taking into consideration the expansion of ${\rm Tr}( \rho)$ as well, expectation values are expanded as 
\begin{align}
    \langle o \rangle & = \langle o \rangle_{\rm GGE} - \beta^{k\prime} \int \dd x\, x  \int_0^1 d\tau \langle q_k^\tau(x), o\rangle^{\rm c}  + \frac{1}{2}  \beta^{k\prime\prime} \int \dd x\, x^2  \int_0^1 d\tau \langle q_k^\tau(x), o\rangle^{\rm c}  \nonumber  \\& +  \beta^{k\prime} \beta^{l\prime} \int \dd x \dd y\, x  y \,  \int_0^1 d\tau \int_0^\tau d\tau' \langle q_k^\tau(x), q_l^{\tau'}(y), o\rangle^{\rm c}. \label{lastline}
\end{align}
By using the symmetry under exchange of indices $k\leftrightarrow l$ on the last line, this indeed is written in terms of the 2- and 3-point KMB connected correlation functions,
\begin{align}
    \langle o \rangle & = \langle o \rangle_{\rm GGE} - \beta^{k\prime} \int \dd x\, x  ( q_k(x), o)  + \frac{1}{2}  \beta^{k\prime\prime} \int \dd x\, x^2  ( q_k(x), o)
    \nonumber  \\& + \frac12 \beta^{k\prime} \beta^{l\prime} \int \dd x \dd y\, x  y \, ( q_k(x), q_l(y), o).
\end{align}

\section{Hydrodynamic expansion calculations} \label{sec:hydroexpcomp}

Let us denote as in the main text
\beq
	\bs U^{(n)} = -\int \dd y\,y^n ( \bs q(0,t),\bs q(y,0)),\quad
	\bs U^{(n,m)} = \int \dd y\,y^n z^m( \bs q(0,t),\bs q(y,0),\bs q(z,0)),
\eeq
and
\beq
	\bs V^{(n)} =  -\int \dd y\,y^n ( \bs j(0,t),\bs q(y,0)),\quad
	\bs V^{(n,m)} = \int \dd y\,y^n z^m( \bs j(0,t),\bs q(y,0),\bs q(z,0)).
\eeq
along with 
\begin{equation}
     U_0^{(n)} = \big[U^{(n)} \big]_{t^0}, \quad  U_0^{(n,m)}=\big[U^{(n,m)} \big]_{t^0}, \quad V_0^{(n)}=\big[V^{(n)} \big]_{t^0}, \quad  V_0^{(n,m)}=\big[V^{(n,m)} \big]_{t^0},
\end{equation}
i.e. their correspondent $O(t^0)= O(1)$ terms in their large $t$ expansion. Recall that these are evaluated in a homogeneous, stationary state characterised by potentials $\beta^i$, or equivalently by the densities $(q_i)$. Moreover, we shall use all the known Euler and diffusive hydrodynamic coefficients in this state: the convective matrix or flux Jacobian $\ms A$ \eqref{Amatrix}, susceptibility matrix $\ms C$ \eqref{CBmatrix}, and diffusion matrix $\mf D$ \eqref{eq:maincoeff1}, \eqref{eq:Onsager}. We also use $\ms C',\,\ms A',\,\mf D'$ as $\ms C_{ijk}' = -\p_{\beta^k} \mathsf C_{ij}$ (eq.~\eqref{C3matrix}, fully symmetric) and $(\ms A')_i^{~jk}  = \p_{(q_k)} \ms A_i^{~j}$ (symmetric in $j,k$), $(\mf D')_i^{~jk} = \p_{(q_k)}\mf D_i^{~j}$. We will also sometimes use $\ms B'_{ijk} = -\p_{\beta^k} \ms B_{ij}$ (symmetric in $j,k$, see \eqref{CBmatrix} for the $\ms B$ matrix). We note that, using $\ms B = \ms A\ms C$,
\beq\label{eq:Bprime}
    \ms B' = \ms A' (\ms C,\ms C) + \ms A\ms C',
\eeq
which we will use to simplify the form of some equations.

We have from \eqref{initj}, and noting that $U^{(0)} = -\ms C$, $U^{(0,0)} = \ms C'$ (here and below, expressions are valid up to, including, 2nd derivatives),
\beqa \label{eq:expansiontot}
	\b{\bs j} &=& ( \bs j) + \bs V^{(1)}\p\bs\beta + \frc12 \bs V^{(2)} \p^2\bs\beta + \frc12 \bs V^{(1,1)} (\p\bs\beta,\p\bs\beta) \n
	{\mathtt q} &=& ( \bs q) + \bs U^{(1)}\p\bs\beta + \frc12 \bs U^{(2)} \p^2\bs\beta + \frc12 \bs U^{(1,1)} (\p\bs\beta,\p\bs\beta) \n
	\p{\mathtt q} &=&  -\ms C\p\beta + U^{(1)}\p^2 \beta + U^{(1,0)}(\p\beta,\p\beta)\n
	\p^2{\mathtt q}&=& -\ms C\p^2\beta + \ms C'(\p\beta,\p\beta).\label{jq}
\eeqa
As mentioned in the main text, to have a self-consistent hydrodynamic equation, we now must express the value of the currents in the first line of \eqref{eq:expansiontot} as functions of the hydrodynamic variables $\b{\bs q}$, rather than the $q \equiv  ( q )$. The relation between the two sets of densities can be inverted as follows:
\begin{equation}\label{qqbar}
   ( q )=\mathtt q-U^{(1)}\p\beta- \frc12 \bs U^{(2)} \p^2\bs\beta - \frc12 \bs U^{(1,1)} (\p\bs\beta,\p\bs\beta).
\end{equation}
Notice how for now, we are keeping $U^{(1)}$, $U^{(2)}$ and $U^{(1,1)}$ evaluated at $(q)$, not $\mathtt q = \b q$. Later in the computation, we will evaluate these at  $\mathtt q$ (that is, integrated correlation functions as per their basic definition, but in the state characterised by $\mathtt q$ instead of $(q)$), which we will denote $\b U^{(1)}$, $\b U^{(2)}$ and $\b U^{(1,1)}$.
Using \eqref{qqbar}, the first term, $(j)$, in the expression for $\b j$ in \eqref{jq}, is written in terms of $\mathtt q$ as follows
\begin{equation}\label{jjbar}
( j ) =  \overline{( j)} - \b{\ms A} U^{(1)}\p\beta -   \frac{1}{2} \b{\ms A} U^{(2)} \p^2\bs\beta  - \frac{1}{2} \b{\ms A} \bs U^{(1,1)} (\p\bs\beta,\p\bs\beta)+ \frc12 \b{\ms A}' (U^{(1)}\p\beta, U^{(1)}\p\beta)
\end{equation}
where $\overline{( j)} = ( j)_{\mathtt q}$ is the current average in the state characterised by $\mathtt q$, and  $\b{\ms A} = \p\overline{( j)}/\p\mathtt q$ is the flux Jacobian in that state. The last term in \eqref{jjbar} comes from the 2nd order term in the Taylor expansion of $(j)_{\mathtt q + \delta}$ in $\delta = -U^{(1)}\p \beta\,+$ higher derivatives.
We then obtain the following expression for the current 
%\beqa
%	\b j &=& \overline{( j)} - (V^{(1)}-\b{\ms A} U^{(1)})\p\beta +
%	\frc12 (V^{(2)} - \ms A U^{(2)})\p^2\beta +
%	\frc12 (V^{(1,1)}-\ms A U^{(1,1)})(\p\beta,\p\beta)
%	+\frc12 \ms A' (U^{(1)}\p\beta, U^{(1)}\p\beta)\n
%	&&-(V^{(1,0)}-AU^{(1,0)}) \ms C  \bs U^{(1)}(\p\bs\beta ,\p\bs\beta )
%	\label{jbar}
%\eeqa
%%
%To be compared with what Ben previously wrote
\begin{align}
    \b j =&\  \overline{( j)} + (V^{(1)}-\b{\ms A} U^{(1)})\p\beta +
	\frc12 (V^{(2)} - \b{\ms A} U^{(2)})\p^2\beta \nonumber \\& +
	\frc12 (V^{(1,1)}-\b{\ms A} U^{(1,1)})(\p\beta,\p\beta)
	+\frc12 \b{\ms A}' (U^{(1)}\p\beta, U^{(1)}\p\beta) \nonumber \\= &\  
 \overline{( j)} + (V^{(1)}-\ms A U^{(1)})\p\beta +
	\frc12 (V^{(2)} - \ms A U^{(2)})\p^2\beta \nonumber \\&  +
	\frc12 (V^{(1,1)}-\ms A U^{(1,1)})(\p\beta,\p\beta)
	-\frc12 \ms A' (U^{(1)}\p\beta, U^{(1)}\p\beta),
	\label{jbar}
\end{align}
where in the second step we have re-expanded all $\b{\ms A}$s and $\b{\ms A}'$ about the state $q$ using $\b {\ms A} = \ms A + \ms A' U^{(1)}\p\beta\, +$ higher derivatives and $\b{\ms A}' = \ms A'\,+$ higher derivatives.

We shall show that hydrodynamic calculations give (see section \ref{hydroexp}), in general, up to vanishing terms in time $o(1)$,
\beq\label{u0}
	\bs U^{(1)} = \ms A\ms C t,\quad
	\bs U^{(2)} = -\ms A^2\ms C t^2 - \mf D \ms C \ |t| + \bs U_{0}^{(2)},
\eeq
\beq\label{u00}
	\bs U^{(1,0)}= -\ms B' t ,
\eeq
and
\beq\label{v0}
	\bs V^{(1)} =  \ms A^2\ms C t + \frc12\mf D\ms C \  \text{sgn}(t),\quad
	\bs V^{(2)} = -\ms A^3\ms C t^2 - (\ms A\mf D + \mf D\ms A) \ms C \ |t| + \bs V_{0}^{(2)},
\eeq
which also allows writing  $\mf L^{(1)}$, defined in \eqref{eq:Onsager1} as $\mf L^{(1)} = \Big[2(V^{1} - \ms A U^{1})\Big]_{t=\infty}$, in the form
\begin{equation}
    \mathfrak{L}^{(1)} = \mf D\ms C,
\end{equation}
the Einstein relation \eqref{einstein}.
% not true, not needed
%\beq
%	\bs V^{0,0} = \ms C',\quad
%	\bs V^{(1,0)}=\bs V^{0,1}= \ms A \ms C' t
%	+ \ms A' (\ms C ,\ms C) t + \bs V_{0}^{1,0}
%\eeq
Note that $U^{(2)}_0,\,V^{(2)}_0$ are independent of $t$ and not determined by the Euler and diffusive scales (the above expansions define them). From this, we find in particular
\beq \label{eq:v0}
    \bs V^{(2)} - \mathsf A \bs U^{(2)} =   -\mf D \ms A \ms C \,|t| + \mf L^{(2)},
\eeq
where we used $\mf L^{(2)} = V^{(2)}_0 - \ms AU^{(2)}_0$ from \eqref{eq:maincoeff2}. Further, we have (see section \ref{hydroexp})
\beq 
    \bs V^{(1,0)} - \mathsf A \bs U^{(1,0)} =
	\ms A'(\ms{AC},\ms{C})t + \frc12(\mf D'
	(\ms C,\ms{C})
	+  \mf D\ms C')\sgn(t),
\eeq
and
\beq\label{eq:w110}
	\bs V^{(1,1)} - \mathsf A \bs U^{(1,1)} =
	\ms A'(\ms{AC},\ms{AC})t^2 + \frc12\mf D'
	(\ms C,\ms{AC})(1 + P)\,|t|
	+ \mf D\ms B'\,|t| + \mathfrak L^{(1,1)},
\eeq
where $P$ is the permutation $P_{~~ij}^{kl} = \delta_i^l\delta_j^k$ and we used $\mf L^{(1,1)} = V^{(1,1)}_0 - \ms AU^{(1,1)}_0$, again from \eqref{eq:maincoeff2}.

We insert these into \eqref{jbar} to obtain
\beqa
	\b j 
	&=& \overline{( j)} + \frc12 \mathfrak{L}^{(1)} \p\beta\,\sgn(t)
	+\frc12 (-\mf D\ms A\ms C |t| + \mf L^{(2)})\p^2\beta\n
	&& +\,\frc 12\Big(
	\ms D'(\ms C,\ms{AC}) |t|
	+\mf D\ms B' |t| +  \mf L^{(1,1)}
	\Big)(\p\beta,\p\beta).
	\label{jbar2}
\eeqa
Notice how the $t^2$ terms have already been cancelled. We now invert the last two equations in \eqref{jq}; we denote by $\b\beta$ the potentials associated to the state characterised by $\mathtt q = \b q$. We have:
\beqa
    -\ms C\p\beta &=&
    \p\mathtt q -\mathsf A\ms C t \p^2\b\beta
    + \ms B't\, (\p\b\beta,\p\b\beta) \n
% \p\mathtt q + \ms At\, \p^2 \mathtt q + \ms A \ms C' t \,(\ms \p\b\beta,\p\b\beta)
%	+ \ms B't\, (\p\b\beta,\p\b\beta) \n
	-\ms C \p^2\beta &=& \p^2 \mathtt q - \ms C'(\p\b\beta,\p\b\beta).
\eeqa
We insert this into \eqref{jbar2} using $\mf L^{(1)} = \mf D\ms C$, together with the expression of the diffusion constant as a function of the $\bar{q}$, i.e., $\mf D = \b{\mf D} - \mf D'U^{(1)} \p\beta = 
	\b{\mf D} - \mf D' \ms{AC}t\p\beta$,
to finally obtain an equation where remarkably, all divergences in time $t$ do simplify, 
\beqa
	\b j
	&=& \overline{( j)}-  \frc12 \b{\mf D}\p\mathtt q \,\sgn(t)
	 +\, \frc12\mf L^{(2)} \p^2 \b \beta  +\,\frc12  \mf L^{(1,1)} (\p\b\beta,
	\p\b\beta),
	\label{jbar4}
\eeqa
and which can be rewritten in terms of the derivative of the hydrodynamic variables $\mathtt q = \bar{q}$, using $\p^2 \mathtt q = \ms C'(\p\b\beta,\p\b\beta) - \ms C \p^2\b\beta$,
\beqa
	\b j &=&  \overline{( j)}-  \frc12 \b{\mf D} \p\mathtt q \,\sgn(t)
	 -\, \frc12\mf L^{(2)}\ms C^{-1} \p^2 \mathtt q
	 +\,\frc12 ( \mf L^{(1,1)} +  \mf L^{(2)} \ms C^{-1} \ms C' )  ( \ms C^{-1} \p\mathtt q,
	\ms C^{-1} \p\mathtt q), \n
\eeqa  
giving eq. \eqref{eq:hydro-expanded} with coefficients defined as in eq. \eqref{eq:maincoeff1}. We stress how it has been important to express all hydrodynamic matrices in terms of $\mathtt q$ in the first derivative order (that is, $\mf {\b D}$ instead of $\mf D$), while in the second order it is sufficient to keep the hydrodynamic matrices as functions of $(q) = \mathtt q + O(\p_x)$, as corrections are of higher derivative order, hence neglected.

\subsection{ Large time expansion of the integrated correlators}\label{hydroexp}

We here shall derive \eqref{u0}-\eqref{eq:w110}. For this purpose, we start with the fundamental expressions of correlation functions of charge-charge and charge-current at the hydrodynamic Euler and diffusive scales, which follow directly from the hydrodynamic equation up to the diffusive scale, and hydrodynamic linear response:
\beq
	S_{qq}(k,t)=\int \dd x\,e^{\ri kx}( q(0,t),q(x,0))
	=\exp\Big[-\ri k\mathsf A t - \frc{k^2}2 \mathfrak D |t|\Big]\mathsf C(k)
\eeq
and
\beq
	S_{jq}(k,t)=\int \dd x\,e^{\ri kx}( j(0,t),q(x,0))
	=\Big(\mathsf A - {\rm sgn}(t) \frc{\ri k}2 \mathfrak D\Big)\exp\Big[-\ri k\mathsf A t - \frc{k^2}2 \mathfrak D |t|\Big]\mathsf C(k).
\eeq
We also define the three-point functions 
\beq
	S_{qqq}(k,t)=\int \dd  x \dd  y\, x y \ e^{\ri kx}(q(0,t), q(x,0),q(y,0)),
\eeq
\beq
	S_{jqq}(k,t)=\int \dd  x \dd  y\, x y\ e^{\ri kx}(j(0,t), q(x,0),q(y,0)).
\eeq
Space-integrated correlators are assumed to be of the form $a_nt^n + a_{n-1}t^{n-1}\sgn(t) + \ldots + a_0 \sgn(t)^n + o(1)$ ($t\to\infty$), and the above expressions are assumed to provide the correct first two leading, positive powers of $t$, that is $a_n$ and $a_{n-1}$.

The quantity $\mathsf C(k)$, above, characterises the initial condition of the hydrodynamic equation for the correlator $(q(0,t),q(x,0))$. In general, it takes the form $\ms C(k) = \mathsf C + \ms C_1 \ri k \sgn(t) + \ms C_2\ (\ri k)^2/2 + \ldots$.

Let us denote ${}' = \p/\p(\ri k)$. Using the general formula 
\begin{equation}
   \p e^{X}/\p \ell\, e^{-X} = \p X/\p\ell + \frc12 [X,\p X/\p\ell] + \ldots ,
\end{equation}
 and denoting $E(k,t) = \exp\Big[-\ri k\mathsf A t - \frc{k^2}2 \mathfrak D |t|\Big]$, we note that
\beq\begin{aligned}
    E'(k,t) =&\ (-\ms At + \ri k\mf D |t|) E(k,t)
    + \frc12 \Big[-\ri k\mathsf A t - \frc{k^2}2 \mathfrak D |t|,-\ms At + \ri k\mf D |t|\Big]E(k,t)
    + \ldots \\ 
    =&\ (-\ms At + \ri k\mf D |t|) E(k,t) + O(k^2).
    \end{aligned}
\eeq
Therefore,
\beq\label{eq:U1app}
	U^{(1)} = -S_{qq}'(0,t)
	= \mathsf A\mathsf Ct  - \ms C_1\sgn(t) + o(1)
\eeq
\beq\label{eq:U2app}
	U^{(2)} = -S_{qq}''(0,t) = -\mathsf A^2 \mathsf Ct^2 - \mathfrak D \mathsf C |t| + 2\ms A \ms C_1 |t| + U^{(2)}_0 + o(1).
\eeq
Note that the existence (finiteness) of $U^{(2)}_0$ in \eqref{eq:U2app} is part of our assumption about the asymptotic expansion at large $t$.

We first show that, by PT symmetry, we must have $\mathsf C_1 = 0$. Indeed, by \eqref{eq:U1app},
\beqa
    2\ms C_1 &=& \int \dd x\,x(( q(x,t),q)
    - ( q(x,-t),q)) - 2\mathsf {AC}t \n
    &=& \int_{-t}^t \dd s\,\Big[
    \int \dd x\,x ( \p_s q(x,s),q) - \mathsf {AC}
    \Big] \n
    &=& \int_{-t}^t \dd s\,\Big[
    \int \dd x\,( j(x,s),q)
     - \mathsf {AC}
    \Big] \n
    &=& \int_{-t}^t \dd s\,\Big[
    \mathsf {AC}
     - \mathsf {AC}
    \Big] \n &=& 0.\label{eq:U1is0}
\eeqa
Thus,
\beq
	U^{(1)} = -S_{qq}'(0,t)
	= \mathsf A\mathsf Ct  + o(1),
\eeq
\beq
	U^{(2)} = -S_{qq}''(0,t) = -\mathsf A^2 \mathsf Ct^2 - \mathfrak D \mathsf C |t| + U^{(2)}_0 + o(1).
\eeq
Similarly, we find
\beq
	V^{(1)} = -S_{jq}'(0,t)
	= \mathsf A^2\mathsf Ct + \frc12 \mathfrak D\mathsf C\,\sgn(t) + o(1),
\eeq
\beq
	V^{(2)} = -S_{jq}''(0,t) = -\mathsf A^3 \mathsf Ct^2 - \mathfrak D \mathsf A \mathsf C |t| - \mathsf A \mathfrak D \mathsf C|t| + V^{(2)}_0 + o(1).
\eeq
This yields \eqref{u0} and \eqref{v0}.

For the three-point functions, more work is needed. We may derive $U^{n,m}$ by solving the diffusive-scale equation satisfied by the 3-point function. Starting with
\beq
	\p_t \mathsf q + \p_x \Big(\mathsf j - \frc12 \mathfrak D \p_x \mathsf q\,\sgn(t)\Big)=0,
\eeq
as an initial value problem from $t=0$, we obtain equations for 2- and 3-point functions by differentiation with respect to $\beta(y,0)$, $\beta(z,0)$. Note how the factor $\sgn(t)$ is necessary for this equation also to represent the hydrodynamic limit of negative-time evolution from $t=0$. Note also that, for general positions in space-time, three-point functions cannot be obtained by such a simple response formalism because of the long-range hydrodynamic correlations that develop from inhomogeneous initial conditions; see \cite{2206.14167}. However, here two of the observables are taken at equal time in the stationary state (at time 0), where no such long-range correlations exist; therefore perturbations of the initial clustering state correctly describe space-time correlations with the third observable at later times.

In the following, we will assume that evolution is towards positive times,
$ t>0$.  It is simple to re-insert the factor $\sgn(t)$, which occurs in conjunction with the diffusion matrix and its derivatives.
Denoting $( qq) := ( q(x,t),q(y,0))$, we obtain
\beq
	\p_t ( qq) + \p_x\Big(
	\mathsf A ( qq) - \frc12 \mathfrak D' (\p_x \mathsf q,( qq))
	-\frc12 \mathfrak D \p_x ( qq)\Big)=0,
\eeq
and by further differentiating and using $ \mathfrak D' (\p_x \mathsf q,( qq)) =  \mathfrak D' P(( qq),\p_x \mathsf q)$ in order to have the right order of implicit indices (where we recall that  $P$ is the permutation $P_{ij}^{~kl} = \delta_i^l\delta_j^k$), for $( qqq) := ( q(x,t),q(y,0),q(z,0))$ and $( qq)_{y\atop z} := ( q(x,t),q({y\atop z},0))$
\beqa
	0&=&\p_t ( qqq) + \p_x\Big(
	\mathsf A' (( qq)_y, ( qq)_z) + \mathsf A( qqq)\n &&
	- \frc12 \mathfrak D'' (\p_x \mathsf q,( qq)_y,( qq)_z)
	- \frc12 \mathfrak D' P(( qq)_y,\p_x ( qq)_z)
	- \frc12 \mathfrak D' (\p_x \mathsf q,( qqq)) \n &&
	-\frc12 \mathfrak D' (\p_x ( qq)_y,( qq)_z)
	-\frc12 \mathfrak D \p_x ( qqq)
	\Big).\label{qqqinho}
\eeqa
We are interested in the steady state, so we obtain
\beqa
	0&=&\p_t ( qqq) - 
	\mathsf A' (\p_y+\p_z)(( qq)_y, ( qq)_z) - \mathsf A(\p_y+\p_z)( qqq)\n &&
	- \frc12 \mathfrak D' P(\p_y+\p_z)\p_z( ( qq)_y,( qq)_z)
	-\frc12 \mathfrak D' (\p_y+\p_z)\p_y(( qq)_y,( qq)_z)\n &&
	-\frc12 \mathfrak D (\p_y+\p_z)^2 ( qqq).
\eeqa
We may now set $x=0$, and we define
\beq
	S_{qqq}(k,k',t)=\int \dd y\dd z\,e^{\ri ky +\ri k'z}( q(0,t),q(y,0),q(z,0)).
\eeq
This satisfies
\beqa
	0&=&\p_t S_{qqq} + 
	(\ri k + \ri k')\mathsf A'  (S_{qq}(k,t), S_{qq}(k',t)) +
	(\ri k + \ri k') \mathsf AS_{qqq}\n &&
	+(k+k')k' \frc12 \mathfrak D'P(S_{qq}(k,t),S_{qq}(k',t))
	+ (k+k')k \frc12 \mathfrak D' (S_{qq}(k,t), S_{qq}(k',t))\n &&
	+ (k+k')^2\frc12 \mathfrak D S_{qqq}.
\eeqa
We may solve this order by order in $k,k'$ to get the various hydrodynamic orders. For now we just want up to order $k,\,k'$. At order 1, we have $S_{qqq}(0,0,t) = \mathsf C'$ (that is, the equation above says that it is constant, and we evaluate the constant by definition of the $\mathsf C'$ tensor). The next order, $O(k,k')$, has the form $(a_1\ri k+b_1\ri k') t + (a_0\ri k+b_0\ri k') \sgn(t) + o(1)$ by our assumption on the large-$t$ asymptotic form, and the solution for $a_1, b_1$ is purely ballistic:
\beq
	S_{qqq}(k,k',t) =
	\mathsf C' - (\ri k + \ri k')\ms B' t
	+ (a_0\ri k + b_0\ri k')\sgn(t)
	+ O(k^2, kk', (k')^2).
\eeq
where we used \eqref{eq:Bprime}.
Note by our assumption, the result is meaningful for the first two powers of $t$, so for $t$ and $t^0$ at the order $O(k,k')$, the $t^0$ therefore having coefficient 0. Therefore
\beq
	U^{(1,0)} = \frc{\p S_{qqq}(k,0,t)}{\p k}\Big|_{k=0}
	= -\ms B't + a_0 \sgn(t) + o(1).
\eeq
By using PT symmetry, we show that $a_0=0$ (much like our proof that $\ms C_1=0$ above):
\beqa
    2a_0 &=& \int \dd y\dd z\,y( (q(0,t),q(y,0),q(z,0))
    -
    (q(0,-t),q(y,0),q(z,0)))
    + 2\ms B' t\n
    &=& \int_{-t}^t \dd s\,\Big[\int \dd y\dd z\,y (\p_s q(0,s),q(y,0),q(z,0))
    + \ms B' \Big]\n
    &=& \int_{-t}^t \dd s\,\Big[-\int \dd y\dd z\, ( j(0,s),q(y,0),q(z,0))
    + \ms B' \Big]\n
    &=& \int_{-t}^t \dd s\,\Big[-\ms B'
    + \ms B' \Big]\n
    &=& 0.
\eeqa
Therefore we conclude
\beq
	U^{(1,0)} 
	= -\ms B't + o(1),
\eeq
which yields \eqref{u00}.

We can similarly evaluate $( jqq)$ by differentiating $\mathsf j - \frc12 \mathfrak D \p_x \mathsf q$ in an inhomogeneous state. This is already done above: it is what's in the big parenthesis on the right-hand side of \eqref{qqqinho}. Taking this in a homogeneous state,
\beqa
	( jqq) &=&
	\mathsf A' (( qq)_y, ( qq)_z) + \mathsf A( qqq)\n &&
	- \frc12 \mathfrak D'P (( qq)_y,\p_x ( qq)_z)
	-\frc12 \mathfrak D' (\p_x ( qq)_y,( qq)_z)
	-\frc12 \mathfrak D \p_x ( qqq).
\eeqa
For the purpose of the derivation, we need $V^{(1,1)} - \mathsf A U^{(1,1)}$, so we evaluate
\beqa
	&& ( jqq) - \mathsf A( qqq) =
	\mathsf A' (( qq)_y, ( qq)_z) \n &&
	- \frc12 \mathfrak D'P (( qq)_y,\p_x ( qq)_z)
	-\frc12 \mathfrak D' (\p_x ( qq)_y,( qq)_z)
	-\frc12 \mathfrak D \p_x ( qqq).
\eeqa
With this, in order to go to order 2 in $k,k'$, we only need the $q$-correlators to order 1 in $k,k'$, which we already have. We get
\beqa
	\lefteqn{S_{jqq}(k,k',t) - \mathsf A S_{qqq}(k,k',t)} && \n
	&=&
	\mathsf A' (S_{qq}(k,t), S_{qq}(k',t)) \n &&
	- \frc12 \mathfrak D' P(S_{qq}(k,t),\ri k' S_{qq}(k',t))
	-\frc12 \mathfrak D' (\ri k S_{qq}(k,t), S_{qq}(k',t))
	-\frc12 \mathfrak D (\ri k+\ri k')S_{qqq}(k,k',t),\n
\eeqa
and we are looking for $V^{(1,0)} - \mathsf A U^{(1,0)} = \p/\p(\ri k)\big|_{k=k'=0}$ and $V^{(1,1)} - \mathsf A U^{(1,1)} = \p^2/\p(\ri k)\p(\ri k')\big|_{k=k'=0}$ of this. Therefore we finally obtain
\beqa
	{V^{(1,0)} - \mathsf A U^{(1,0)}} = -
	\mathsf A' (U^{(1)}, \mathsf C) 
	- \frc12 \mathfrak D' (\mathsf C,\mathsf C)
	-\frc12 \mathfrak D \mathsf C',
\eeqa
and
\begin{align}
  &  V^{(1,1)} - \mathsf A U^{(1,1)} =  \mathsf A' (U^{(1)}, U^{(1)}) 
	+\frc12 \mathfrak D' P(U^{(1)},\mathsf C)
	+ \frc12 \mathfrak D' (\mathsf C,U^{(1)})
	+ \mf D \ms B' t + O(1),
\end{align}
which demonstrates \eqref{eq:w110}.

\section{Hydrodynamic coefficients in the free fermions theory}\label{sec:FFfreeexp}
We here explicitly compute the hydrodynamic coefficients in the free fermionic theory introduced in sec. \ref{sec:FreeF}. We start with the computation of the two-point functions on a stationary state is given by momentum occupation function $n(k)$
\begin{align}
    [U^{(2)}]_{ki} &  = - \int d\tau \int \dd x x^2 (q_k(x,\tau),  q_i(t)) \nonumber \\&
    =  -\int_{k_1,k_2} \frac{1}{2\pi} n(k_1) (1-n(k_2)) \int \dd x x^2 \int d\tau e^{- i x (k_1- k_2)} e^{\tau (w(k_2) - w(k_1))} e^{it  (\varepsilon(k_2) - \varepsilon(k_1))}  \nonumber  \\& \times \ms h_i((k_1+k_2)/2) \ms h_k((k_1+k_2)/2)  , 
\end{align}

\begin{align}
    [V^{(2)}]_{ki} &   
    =  \int_{k_1,k_2} \frac{1}{2\pi}  n(k_1) (1-n(k_2)) \int \dd x x^2 \int d\tau e^{- i x (k_1- k_2)} e^{\tau (w(k_2) - w(k_1))}  \nonumber  \\& \times \ms h_i((k_1+k_2)/2) \ms h_k((k_1+k_2)/2) 
 \frac{\varepsilon(k_1) -\varepsilon(k_2)}{k_1- k_2} e^{it  (\varepsilon(k_2) - \varepsilon(k_1))}.
\end{align}
The KMS imaginary time integration gives 
\begin{equation}
    \int_\tau e^{\tau (w(k_2) - w(k_1))} = \frac{1 - e^{w(k_2) - w(k_1)}}{w(k_1) - w(k_2)} = G^{[1]}(k_1,k_2),
\end{equation}
where the function $w(k) = \beta^i \ms h_i(k) = \log(1/n(k) -1)$ characterises the stationary state with occupation $n(k)$. 
Therefore, neglecting the term where derivatives act on the energy phases (which gives the terms of order $t^2$ in the correlator \eqref{u0}, \eqref{v0}) and using $\ms A_{ij} = \int \dd k v(k) \ms h_i(k) \ms h_j(k)$ and $\sum_i \ms h_i(k) \ms h_i(k') = \delta(k-k')$ we obtain 
\begin{align}
    &  \mathfrak{L}^{(2)}   =  \Big[\int \dd k_1 \dd k_2\frac{1}{2\pi}   \delta''(k_1-k_2) n(k_1) (1- n(k_2)) G^{[1]}(k_1,k_2) \nonumber \\& \times \ms h_i((k_1 + k_2)/2) \ms h_k((k_1 + k_2)/2)  \frac{\varepsilon(k_1) -\varepsilon(k_2) }{k_1 - k_2} \nonumber \\& 
    -  \int \dd k_1 \dd k_2  \frac{1}{2\pi} 
 \delta''(k_1-k_2) n(k_1) (1- n(k_2)) G^{[1]}(k_1,k_2) \nonumber \\& \times \ms h_i((k_1 + k_2)/2) \ms h_k((k_1 + k_2)/2)  v((k_1 + k_2)/2)\Big].
\end{align}
Since the measure of integration  $n(k_1) (1- n(k_2)) G^{[1]}(k_1,k_2)$ is symmetric under exchange of $k_1$ and $k_2$, the only term surviving the difference is when the derivative acts on the velocity part, in both terms, therefore, using 
\begin{equation}
    \frac{ (\varepsilon(k_1) -  \varepsilon(k_2))}{k_1 - k_2} - v((k_1+ k_2)/2)  = [1/6 \varepsilon''' - 1/8 \varepsilon''' ](k_1- k_2)^2 + \ldots.
\end{equation}
We finally obtain
\begin{align}
    & \mathfrak{L}^{(2)} =  \int \frac{\dd k}{2 \pi } n(k) (1-n(k) \  \ms  h_i(k) \ms h_k(k) [ \frac{1}{12} v''(k)].
\end{align}
 Notice the importance of having form factors  $\ms  h_i$ which are functions only of $k_1+k_2$ up to possible deviations of orders $(k_1-k_2)^3$, namely, even if in this computation here we have chosen charge densities in the fully symmetric gauge, the result only depends on imposing 1st and $2-$nd order gauge.

 We then move to the computation of the 3-point function.
We first rewrite it as follows 
\begin{equation}
   [U^{(1,1)}]_{kli} = \int \dd x \dd y xy (q_i(t), q_k(x), q_l(y) ) + (k \leftrightarrow l) ,
\end{equation}
\begin{equation}
   [V^{(1,1)}]_{kli} = \int \dd x \dd y xy (j_i(t), q_k(x), q_l(y) ) + (k \leftrightarrow l)  ,
\end{equation}
where we also have used symmetry under exchanges of all indices. We then decompose it as 
\begin{align}
   &    {U}_{ikl}  = (2 \pi)^{-5} \int d\tau d\tau' \int \dd x \dd y x y \int_{k_1,k_2,k_3,k_4,k_5,k_6}  e^{- i x (k_3- k_4)} e^{- i y (k_5- k_6)}   e^{it  (\varepsilon(k_2) - \varepsilon(k_1))}  \nonumber \\& \times \ms 
 h_i((k_1+k_2)/2)  \ms h_k((k_3+k_4)/2)  \ms h_l((k_5+k_6)/2)  \langle c^\dagger_{k_1} c_{k_2} c^\dagger_{k_3} c_{k_4}  c^\dagger_{k_5} c_{k_6} \rangle^c  \nonumber \\& \times  e^{-\tau (w(k_1) - w(k_2))}  e^{-\tau' (w(k_3) - w(k_4))}\nonumber  \\& 
    = (2 \pi)^{-3}\int d\tau d\tau' \int \dd x \dd y x y \int_{k_1,k_2,k_3} n(k_1)(1-n(k_2)) n(k_3)   e^{- i x (k_3- k_1)} e^{- i y (k_2- k_3)} e^{it  (\varepsilon(k_2) - \varepsilon(k_1))}   \nonumber \\&  \times \ms h_i((k_1+k_2)/2)   \ms 
 h_k((k_3+k_1)/2)  \ms  h_l((k_2+k_3)/2) \nonumber  \\ &  
    +(2 \pi)^{-3} \int d\tau d\tau' \int \dd x \dd y x y \int_{k_1,k_2,k_3} n(k_1)(1-n(k_2)) (1-n(k_3))  e^{it  (\varepsilon(k_2) - \varepsilon(k_1))}   e^{- i x (k_2- k_3)}  \nonumber \\ & \times e^{- i y (k_3- k_1)}  e^{-\tau (w(k_1) - w(k_2))}  e^{-\tau' (w(k_2) - w(k_3))}   \ms 
 h_i((k_1+k_2)/2)  \ms 
 h_k((k_2+k_3)/2) \ms  h_l((k_3+k_1)/2) .
\end{align}
Notice how the correlator splits into a contribution where either particle excitations on the two states are the same or where hole expectations are the same. 
The KMS time integration now it gives 
\begin{equation}
    \int_{\tau, \tau'} 
e^{-\tau (w(k_1) - w(k_2))}  e^{-\tau' (w(k_3) - w(k_1))} = \frac{ \frac{1 - e^{w(k_2) - w(k_1)}}{ w(k_1) - w(k_2)} - \frac{1 - e^{w(k_2) - w(k_3)}}{ w(k_3) - w(k_2)} }{w(k_3) -w(k_1)},
\end{equation}
\begin{equation}
    \int_{\tau, \tau'}  e^{-\tau (w(k_1) - w(k_2))}  e^{-\tau' (w(k_2) - w(k_3))} =  \frac{ \frac{1 - e^{w(k_2) - w(k_1)}}{ w(k_1) - w(k_2)} - \frac{1 - e^{w(k_3) - w(k_1)}}{ w(k_1) - w(k_3)} }{w(k_2) -w(k_3)}.
\end{equation}
We are here interested in the limits of coinciding momenta as 
\begin{equation}
    \lim_{k_3 \to k_1} \frac{ \frac{1 - e^{w(k_2) - w(k_1)}}{ w(k_1) - w(k_2)} - \frac{1 - e^{w(k_2) - w(k_3)}}{ w(k_3) - w(k_2)} }{w(k_3) -w(k_1)} =- \frac{\partial G^{[1]}(k_1,k_2)}{\partial w(k_1)} = \frac{\partial G^{[1]}(k_1,k_2)}{\partial w(k_2)},
\end{equation}
\begin{equation}
    \lim_{k_3 \to k_2} \frac{ \frac{1 - e^{w(k_2) - w(k_1)}}{ w(k_1) - w(k_2)} - \frac{1 - e^{w(k_3) - w(k_1)}}{ w(k_1) - w(k_3)} }{w(k_2) -w(k_3)} = -\frac{\partial G^{[1]}(k_1,k_2)}{\partial w(k_1)} = \frac{\partial G^{[1]}(k_1,k_2)}{\partial w(k_2)},
\end{equation}
given that integration over spaces is such as 
\begin{equation}
   \int \dd x \dd y \  xy  \ e^{- i x (k_3- k_1)} e^{- i y (k_2- k_3)}= - (2 \pi)^2 \delta'(k_3- k_1) \delta'(k_2- k_3)  ,
\end{equation}
\begin{equation}
   \int \dd x \dd y \  xy  \ e^{- i x (k_2- k_3)} e^{- i y (k_3- k_1)}   = - (2 \pi)^2 \delta'(k_2- k_3) \delta'(k_3- k_1)  .
\end{equation}
As before for the case of 2 point functions, since we have $ [ \frac{\varepsilon(k_1) - \varepsilon(k_2)}{k_1- k_2}    - v((k_1 + k_2)/2)]  = (k_1 - k_2)^2 \varepsilon'''/24 + \ldots$ the derivative of the second delta function can only act on the first delta to produce a second order derivative in momentum, all other terms contribute to zero when taking the difference $V^{(1,1)} - A U^{(1,1)}$, giving therefore 
\begin{equation}
   \mathfrak L^{(1,1)} =  -   \int \frac{\dd k}{2\pi} \  n(k) (1-n(k))(2 n(k) -1 ) \ms h_i(k) \ms h_l(k) [ \frac{1}{12} v''(k)] \ \ms h_k(k),
\end{equation}
 where we have 
\begin{equation}
    n(k) (1-n(k))[ 2 n(k) -1 ] \ms h_k(k)  = \frac{ \delta  (n (1- n))}{ \delta {\beta^k} } ,
\end{equation}
which yields eq. \eqref{eq:zzfree}.

\section{Hydrodynamic coefficients in  interacting quantum integrable models under symmetric gauge}\label{sec:FFfreeexp2}
Under the symmetric gauge, the computation of $\mf L^{(2)}$ in the interacting case follows the free fermion case directly, see sec. \ref{sec:FFfreeexp}.

The above form factors give the following exact  expansion
\begin{equation}
U^{(2)}  =  t^2 U^{{(2)}}_{\rm 1ph-(1)}  +     t U^{{(2)}}_{\rm 2ph} +  U^{{(2)}}_{\rm 1ph-(2)} ,
\end{equation}
\begin{equation}
V^{(2)} =   t^2 V^{{(2)}}_{\rm 1ph-(1)} +     t V^{{(2)}}_{\rm 2ph} +  V^{{(2)}}_{\rm 1ph-(2)}.
\end{equation}
Namely for each correlator, there are two contributions, one of order $t^2$ and one of order $t^0$ coming from the one particle-hole contributions, and one from the two particle-hole contributions, and nothing else. 
The terms with 2ph give the diffusive terms, namely all terms in the expansion of the integrated correlator, see eq. \eqref{v0} and \eqref{eq:w110} which are proportional to the diffusion matrix $\mf D$. We therefore here focus on the one particle-hole contributions. 
Denoting with $F^{(q_i)}(p,h) $ the one particle-hole form factors of the density $q_i(0)$, we have the two correlators entering   
\begin{align}
    [U^{{(2)}}_{\rm 1ph-(2)}  ]_{ki} & 
    =  \int_{p,h} {k'(p)}{} \rho(h) (1-n(p)) G^{[1]}(p,h) F^{(q_k)}(p,h)   F^{(q_i)}(p,h)   \delta''(k(p) - k(h)), 
\end{align}
\begin{align}
    [V^{{(2)}}_{\rm 1ph - (2)} ]_{ki} & 
    =  \int_{p,h} {k'(p)}{} \rho(h) (1-n(p)) G^{[1]}(p,h)\frac{\varepsilon(p) - \varepsilon(h)}{k(p)-k(h)} \\&  \times  F^{(q_k)}(p,h) F^{(q_i)}(p,h)    \delta''(k(p) - k(h)) ,
\end{align}
and where we introduced the KMS imaginary time integration  
\begin{equation}
    \int_\tau e^{\tau (w^{\rm dr}(p) - w^{\rm dr}(h))} =  G^{[1]}(p,h).
\end{equation}
Working in the 2nd order symmetric gauge imposes that one particle-hole form factors cannot depend on momentum difference $k(p)-k(h)$ up to deviations $O((k(p)-k(h))^3)$. Knowing the well-established relation between form factors and single-particle eigenvalues $h_i(h)$ of charge density $q_i$
\begin{equation}
   F^{(q_i)}(h,h) = \ms h^{\rm dr}_i(h) , \quad \quad  h^{\rm dr}_i= (1- T n)^{-1}h_i,
\end{equation}
and using that momentum $k(\theta)$ is a monotonous function in $\theta$, this implies  
\begin{equation}
    F^{(q_i)}(p,h)  = \ms h^{\rm dr}_i((p+h)/2) +O((k(p)-k(h))^3). 
\end{equation}
Therefore,  using $\ms A_{ij} =  (\ms h_i , (1- n T)^{-1} v^{\rm eff} (1-n T), \ms h_j)$, $\ms h_i^{\rm dr} = (1 - T n)^{-1} \ms h_i$ and $\sum_i \ms h_i(\theta) \ms h_i(\theta') = \delta(\theta-\theta')$, we finally obtain 
\begin{align}
    &  \mathfrak{L}^{(2)}   = \int_{p,h} k'(p) \rho(h) (1-n(p)) G^{[1]}(p,h)  \ms h^{\rm dr}_i((p+h)/2)   \ms h^{\rm dr}_k((p+h)/2) \nonumber \\&   \delta''(k(p) - k(h)) \Big[ \frac{\varepsilon(p) - \varepsilon(h)}{k(p)-k(h)}  - v^{\rm eff} ((h+p)/2) +O((k(p)-k(h))^3) \Big] .
\end{align}
Choosing $p=h+\epsilon$ we can integrate over $\delta$ and the derivative must act on the parenthesis, which is indeed of order $\epsilon^2$
\begin{equation}
 \Big[ \frac{\varepsilon(p) - \varepsilon(h)}{k(p)-k(h)}  - v^{\rm eff} ((h+p)/2) \Big]  = \epsilon^2 \frac{ \varepsilon'''(h) k'(h) - \varepsilon'(h) k'''(h) }{24 (k'(h))^2 } + \ldots
\end{equation}
We obtain then, using $\delta''(k(p)-k(h)) = \delta''(\epsilon k'(h))$ and $\varepsilon'(\theta) = v^{\rm eff}(\theta) k'(\theta)$
\begin{align}
    &  \mathfrak{L}^{(2)}   =\int_{\theta}  \rho(\theta) (1-n(\theta))  \ms h^{\rm dr}_i(\theta) \ms h^{\rm dr}_k(\theta) 
 \frac{k' (\theta)\varepsilon'''(\theta) - k'''(\theta) \varepsilon'(\theta)}{12(k'(\theta))^4} ,
\end{align}
where we used that $G^{[1]}(h,h)=1$. The computation of the three-point function follows analogously, as in the free fermions case, see sec. \ref{sec:FFfreeexp}. 

\section{The centre of mass operator and the symmetric gauge in free fermions}\label{sec:freeFGauge}
The free fermionic charges reported in eq. \eqref{eq:chargesFF} 
\begin{equation} 
    q_i(0)= \int \dd y  \int \frac{\dd k}{2 \pi} \mathsf h_i(k) e^{i y k} c^\dagger_{-y/2} c_{y/2} =  \int \frac{\dd k_1}{2 \pi} \frac{\dd k_2}{2 \pi}   \mathsf h_i((k_1 + k_2)/2) c^\dagger_{k_1} c_{k_2},
\end{equation}
are directly expressed in the fully symmetric gauge, as their matrix elements $\mathsf h_i((k_1 + k_2)/2)$ are only functions of the sum of the two momenta. In general, one could choose any charge with the following expression 
\begin{equation} 
    \tilde{q}=   \int \frac{\dd k_1}{2 \pi} \frac{\dd k_2}{2 \pi}    \tilde{\mathsf h}( k_2,k_1) c^\dagger_{k_1} c_{k_2},
\end{equation}
with a generic matrix element, function of the two momenta $ \mathsf h( k_2,k_1)$.
The centre of mass operator $X$ necessary to introduce the gauge shifts is given in free fermion theories by 
\begin{equation}
    X  =- \int \dd x \ x \ c^\dagger_x c_x = - \int \frac{\dd k_1 \dd k_2}{2 \pi}    \ii \delta'(k_1- k_2)  c^\dagger_{k_1} c_{k_2}.
\end{equation}
We can consider the matrix element of first gauge shift $a^{(1)}_i =  \frac{1}{2}  \{ X,   \tilde{q} \}$ between the thermodynamic state $\langle n |$ with filling $n(k)$ and its one particle-hole excitation $| n, \{ {h},{p} \} \rangle$ with momenta $p$ and $h$: by explicit computation we obtain 
\begin{align}
\langle n |     \tilde{q} X | n, \{ {h},{p} \} \rangle &  = - \int \frac{\dd k_1}{2 \pi} \dd k_2  
  \int \frac{\dd q}{2 \pi} \frac{\dd k}{2 \pi}     \tilde{\mathsf h}( q,k) \ii \delta'(k_1- k_2)   \langle n |  c^\dagger_{k_1} c_{k_2} c^\dagger_{k} c_{q} | n, \{ {h},{p} \} \rangle  \nonumber \\&   =  \ii \partial_{p-h}  \tilde{\mathsf h}( p,h)
\end{align}
Therefore we conclude that the new charge density, after the shift $ \tilde{q} \to  \tilde{q} +\partial_x a^{(1)}_i$, has its matrix elements given by 
\begin{equation}
   \tilde{\mathsf h}( k_2,k_1) \to   \tilde{\mathsf h}( k_2,k_1) -   (k_2- k_1) \partial_{k_2-k_1} \tilde{\mathsf h}( k_2,k_1) .
\end{equation}
The higher order relations, necessary to cancel $k_2- k_1$ dependence at higher orders, then follows from repeating a similar argument with higher powers of the operator $X$.
\bibliographystyle{ieeetr.bst}
\bibliography{biblio.bib}

\begin{thebibliography}{10}

\bibitem{ldlandau2013}
L.~D. Landau, {\em Fluid Mechanics: Landau and Lifshitz: Course of Theoretical
  Physics, Volume 6}.
\newblock Pergamon, sep 2013.

\bibitem{Rangamani2009}
M.~Rangamani, ``Gravity and hydrodynamics: lectures on the fluid-gravity
  correspondence,'' {\em Classical and Quantum Gravity}, vol.~26, p.~224003,
  Oct. 2009.

\bibitem{deBoer2015}
J.~de~Boer, M.~P. Heller, and N.~Pinzani-Fokeeva, ``Effective actions for
  relativistic fluids from holography,'' {\em Journal of High Energy Physics},
  vol.~2015, Aug. 2015.

\bibitem{Lucas2015}
A.~Lucas, ``Hydrodynamic transport in strongly coupled disordered quantum field
  theories,'' {\em New Journal of Physics}, vol.~17, p.~113007, Oct. 2015.

\bibitem{Lucas2018}
A.~Lucas and K.~C. Fong, ``Hydrodynamics of electrons in graphene,'' {\em
  Journal of Physics: Condensed Matter}, vol.~30, p.~053001, Jan. 2018.

\bibitem{Ku2020}
M.~J.~H. Ku, T.~X. Zhou, Q.~Li, Y.~J. Shin, J.~K. Shi, C.~Burch, L.~E.
  Anderson, A.~T. Pierce, Y.~Xie, A.~Hamo, U.~Vool, H.~Zhang, F.~Casola,
  T.~Taniguchi, K.~Watanabe, M.~M. Fogler, P.~Kim, A.~Yacoby, and R.~L.
  Walsworth, ``Imaging viscous flow of the dirac fluid in graphene,'' {\em
  Nature}, vol.~583, pp.~537--541, July 2020.

\bibitem{PhysRevLett.122.090601}
M.~Schemmer, I.~Bouchoule, B.~Doyon, and J.~Dubail, ``Generalized hydrodynamics
  on an atom chip,'' {\em Phys. Rev. Lett.}, vol.~122, p.~090601, Mar 2019.

\bibitem{2009.06651}
N.~Malvania, Y.~Zhang, Y.~Le, J.~Dubail, M.~Rigol, and D.~S. Weiss,
  ``Generalized hydrodynamics in strongly interacting 1d bose gases,'' 2020.
\newblock arXiv:2009.06651.

\bibitem{2006.08577}
F.~Møller, C.~Li, I.~Mazets, H.-P. Stimming, T.~Zhou, Z.~Zhu, X.~Chen, and
  J.~Schmiedmayer, ``Extension of the generalized hydrodynamics to dimensional
  crossover regime,'' 2020.
\newblock arXiv:2006.08577.

\bibitem{2009.13535}
A.~Scheie, N.~E. Sherman, M.~Dupont, S.~E. Nagler, M.~B. Stone, G.~E. Granroth,
  J.~E. Moore, and D.~A. Tennant, ``Detection of kardar-parisi-zhang
  hydrodynamics in a quantum heisenberg spin-$1/2$ chain,'' 2020.
\newblock arXiv:2009.13535.

\bibitem{PhysRevLett.128.021604}
E.~Granet, B.~Bertini, and F.~H.~L. Essler, ``Duality between weak and strong
  interactions in quantum gases,'' {\em Phys. Rev. Lett.}, vol.~128, p.~021604,
  Jan 2022.

\bibitem{Crossley2017}
M.~Crossley, P.~Glorioso, and H.~Liu, ``Effective field theory of dissipative
  fluids,'' {\em Journal of High Energy Physics}, vol.~2017, Sept. 2017.

\bibitem{PhysRevD.85.085029}
S.~Dubovsky, L.~Hui, A.~Nicolis, and D.~T. Son, ``Effective field theory for
  hydrodynamics: Thermodynamics, and the derivative expansion,'' {\em Phys.
  Rev. D}, vol.~85, p.~085029, Apr 2012.

\bibitem{10.21468/SciPostPhys.12.4.130}
S.~Pappalardi, L.~Foini, and J.~Kurchan, ``{Quantum bounds and
  fluctuation-dissipation relations},'' {\em SciPost Phys.}, vol.~12, p.~130,
  2022.

\bibitem{PhysRevLett.122.251601}
S.~c.~v. Grozdanov, P.~K. Kovtun, A.~O. Starinets, and
  P.~Tadi\ifmmode~\acute{c}\else \'{c}\fi{}, ``Convergence of the gradient
  expansion in hydrodynamics,'' {\em Phys. Rev. Lett.}, vol.~122, p.~251601,
  Jun 2019.

\bibitem{PhysRevLett.118.226601}
T.~Scaffidi, N.~Nandi, B.~Schmidt, A.~P. Mackenzie, and J.~E. Moore,
  ``Hydrodynamic electron flow and hall viscosity,'' {\em Phys. Rev. Lett.},
  vol.~118, p.~226601, Jun 2017.

\bibitem{PhysRevX.6.041065}
O.~A. Castro-Alvaredo, B.~Doyon, and T.~Yoshimura, ``Emergent hydrodynamics in
  integrable quantum systems out of equilibrium,'' {\em Phys. Rev. X}, vol.~6,
  p.~041065, Dec 2016.

\bibitem{PhysRevLett.117.207201}
B.~Bertini, M.~Collura, J.~{De Nardis}, and M.~Fagotti, ``Transport in
  out-of-equilibrium $xxz$ chains: Exact profiles of charges and currents,''
  {\em Phys. Rev. Lett.}, vol.~117, p.~207201, Nov 2016.

\bibitem{SciPostPhys.2.2.014}
B.~Doyon and T.~Yoshimura, ``{A note on generalized hydrodynamics:
  inhomogeneous fields and other concepts},'' {\em SciPost Phys.}, vol.~2,
  p.~014, 2017.

\bibitem{PhysRevLett.125.240604}
A.~Bastianello, A.~De~Luca, B.~Doyon, and J.~{De Nardis}, ``Thermalization of a
  trapped one-dimensional bose gas via diffusion,'' {\em Phys. Rev. Lett.},
  vol.~125, p.~240604, Dec 2020.

\bibitem{vir1}
V.~B. Bulchandani, R.~Vasseur, C.~Karrasch, and J.~E. Moore, ``Solvable
  hydrodynamics of quantum integrable systems,'' {\em Phys. Rev. Lett.},
  vol.~119, p.~220604, Nov 2017.

\bibitem{PhysRevB.101.180302}
A.~J. Friedman, S.~Gopalakrishnan, and R.~Vasseur, ``Diffusive hydrodynamics
  from integrability breaking,'' {\em Phys. Rev. B}, vol.~101, p.~180302, May
  2020.

\bibitem{2005.13546}
J.~Lopez-Piqueres, B.~Ware, S.~Gopalakrishnan, and R.~Vasseur, ``Hydrodynamics
  of non-integrable systems from relaxation-time approximation,'' 2020.
\newblock arXiv:2005.13546.

\bibitem{Gopalakrishnan2019}
S.~Gopalakrishnan, R.~Vasseur, and B.~Ware, ``Anomalous relaxation and the
  high-temperature structure factor of {XXZ} spin chains,'' {\em Proceedings of
  the National Academy of Sciences}, vol.~116, pp.~16250--16255, July 2019.

\bibitem{PhysRevB.102.115121}
M.~Fava, B.~Ware, S.~Gopalakrishnan, R.~Vasseur, and S.~A. Parameswaran, ``Spin
  crossovers and superdiffusion in the one-dimensional hubbard model,'' {\em
  Phys. Rev. B}, vol.~102, p.~115121, Sep 2020.

\bibitem{PhysRevB.96.081118}
E.~Ilievski and J.~De~Nardis, ``Ballistic transport in the one-dimensional
  hubbard model: The hydrodynamic approach,'' {\em Phys. Rev. B}, vol.~96,
  p.~081118, Aug 2017.

\bibitem{PhysRevLett.125.070601}
J.~{De Nardis}, S.~Gopalakrishnan, E.~Ilievski, and R.~Vasseur,
  ``Superdiffusion from emergent classical solitons in quantum spin chains,''
  {\em Phys. Rev. Lett.}, vol.~125, p.~070601, Aug 2020.

\bibitem{PhysRevLett.124.210605}
J.~{De Nardis}, M.~Medenjak, C.~Karrasch, and E.~Ilievski, ``Universality
  classes of spin transport in one-dimensional isotropic magnets: The onset of
  logarithmic anomalies,'' {\em Phys. Rev. Lett.}, vol.~124, p.~210605, May
  2020.

\bibitem{PhysRevLett.124.140603}
P.~Ruggiero, P.~Calabrese, B.~Doyon, and J.~Dubail, ``Quantum generalized
  hydrodynamics,'' {\em Phys. Rev. Lett.}, vol.~124, p.~140603, Apr 2020.

\bibitem{Bulchandani_2019}
V.~B. Bulchandani, X.~Cao, and J.~E. Moore, ``Kinetic theory of quantum and
  classical toda lattices,'' {\em Journal of Physics A: Mathematical and
  Theoretical}, vol.~52, p.~33LT01, jul 2019.

\bibitem{SciPostPhys.3.6.039}
B.~Doyon and H.~Spohn, ``{Drude Weight for the Lieb-Liniger Bose Gas},'' {\em
  SciPost Phys.}, vol.~3, p.~039, 2017.

\bibitem{Doyon_2017}
B.~Doyon and H.~Spohn, ``Dynamics of hard rods with initial domain wall
  state,'' {\em Journal of Statistical Mechanics: Theory and Experiment},
  vol.~2017, p.~073210, jul 2017.

\bibitem{PhysRevB.96.220302}
M.~Fagotti, ``Higher-order generalized hydrodynamics in one dimension: The
  noninteracting test,'' {\em Phys. Rev. B}, vol.~96, p.~220302, Dec 2017.

\bibitem{10.21468/SciPostPhys.8.3.048}
M.~Fagotti, ``{Locally quasi-stationary states in noninteracting spin
  chains},'' {\em SciPost Phys.}, vol.~8, p.~048, 2020.

\bibitem{PhysRevLett.128.190401}
B.~Bertini, F.~H.~L. Essler, and E.~Granet,
  ``Bogoliubov-born-green-kirkwood-yvon hierarchy and generalized
  hydrodynamics,'' {\em Phys. Rev. Lett.}, vol.~128, p.~190401, May 2022.

\bibitem{1912.08496}
B.~Doyon, ``Lecture notes on generalised hydrodynamics,'' 2019.
\newblock arXiv:1912.08496.

\bibitem{doyon_hydrodynamic_2022}
B.~Doyon, ``Hydrodynamic {Projections} and the {Emergence} of {Linearised}
  {Euler} {Equations} in {One}-{Dimensional} {Isolated} {Systems},'' {\em
  Communications in Mathematical Physics}, vol.~391, pp.~293--356, Apr. 2022.

\bibitem{10.21468/SciPostPhys.6.4.049}
J.~{De Nardis}, D.~Bernard, and B.~Doyon, ``{Diffusion in generalized
  hydrodynamics and quasiparticle scattering},'' {\em SciPost Phys.}, vol.~6,
  p.~49, 2019.

\bibitem{Durnin2021}
J.~Durnin, A.~D. Luca, J.~D. Nardis, and B.~Doyon, ``Diffusive hydrodynamics of
  inhomogenous hamiltonians,'' {\em Journal of Physics A: Mathematical and
  Theoretical}, vol.~54, p.~494001, Nov. 2021.

\bibitem{Casati}
R.~Luo, G.~Benenti, G.~Casati, and J.~Wang, ``Onsager reciprocal relations with
  broken time-reversal symmetry,'' {\em Phys. Rev. Research}, vol.~2,
  p.~022009, Apr 2020.

\bibitem{PhysRevE.103.052116}
S.~Giordano, ``Entropy production and onsager reciprocal relations describing
  the relaxation to equilibrium in stochastic thermodynamics,'' {\em Phys. Rev.
  E}, vol.~103, p.~052116, May 2021.

\bibitem{PhysRevE.102.030101}
A.~Coretti, L.~Rondoni, and S.~Bonella, ``Fluctuation relations for systems in
  a constant magnetic field,'' {\em Phys. Rev. E}, vol.~102, p.~030101, Sep
  2020.

\bibitem{2208.01463}
Z.~Krajnik, J.~Schmidt, V.~Pasquier, T.~Prosen, and E.~Ilievski, ``Universal
  anomalous fluctuations in charged single-file systems,'' 2022.

\bibitem{Popkov2015}
V.~Popkov, A.~Schadschneider, J.~Schmidt, and G.~M. Sch\"{u}tz, ``Fibonacci
  family of dynamical universality classes,'' {\em Proceedings of the National
  Academy of Sciences}, vol.~112, pp.~12645--12650, Sept. 2015.

\bibitem{2103.01976}
V.~B. Bulchandani, S.~Gopalakrishnan, and E.~Ilievski, ``Superdiffusion in spin
  chains,'' 2021.
\newblock arXiv:2103.01976.

\bibitem{Jara2015}
M.~Jara, T.~Komorowski, and S.~Olla, ``Superdiffusion of energy in a chain of
  harmonic oscillators with noise,'' {\em Communications in Mathematical
  Physics}, vol.~339, pp.~407--453, July 2015.

\bibitem{Scheie2021}
A.~Scheie, N.~E. Sherman, M.~Dupont, S.~E. Nagler, M.~B. Stone, G.~E. Granroth,
  J.~E. Moore, and D.~A. Tennant, ``Detection of
  kardar{\textendash}parisi{\textendash}zhang hydrodynamics in a quantum
  heisenberg spin-1/2 chain,'' {\em Nature Physics}, Mar. 2021.

\bibitem{Bulchandani2020}
V.~B. Bulchandani, C.~Karrasch, and J.~E. Moore, ``Superdiffusive transport of
  energy in one-dimensional metals,'' {\em Proceedings of the National Academy
  of Sciences}, vol.~117, pp.~12713--12718, May 2020.

\bibitem{Spohn2014}
H.~Spohn, ``Nonlinear fluctuating hydrodynamics for anharmonic chains,'' {\em
  Journal of Statistical Physics}, vol.~154, pp.~1191--1227, Feb. 2014.

\bibitem{10.21468/SciPostPhys.9.5.075}
M.~Medenjak, J.~D. Nardis, and T.~Yoshimura, ``{Diffusion from Convection},''
  {\em SciPost Phys.}, vol.~9, p.~75, 2020.

\bibitem{Doyon2021}
B.~Doyon and J.~Durnin, ``Free energy fluxes and the
  kubo{\textendash}martin{\textendash}schwinger relation,'' {\em Journal of
  Statistical Mechanics: Theory and Experiment}, vol.~2021, p.~043206, Apr.
  2021.

\bibitem{Wan2006}
W.~Wan, S.~Jia, and J.~W. Fleischer, ``Dispersive superfluid-like shock waves
  in nonlinear optics,'' {\em Nature Physics}, vol.~3, pp.~46--51, Dec. 2006.

\bibitem{PhysRevLett.101.170404}
J.~J. Chang, P.~Engels, and M.~A. Hoefer, ``Formation of dispersive shock waves
  by merging and splitting bose-einstein condensates,'' {\em Phys. Rev. Lett.},
  vol.~101, p.~170404, Oct 2008.

\bibitem{PhysRevA.85.033603}
A.~M. Kamchatnov and N.~Pavloff, ``Generation of dispersive shock waves by the
  flow of a bose-einstein condensate past a narrow obstacle,'' {\em Phys. Rev.
  A}, vol.~85, p.~033603, Mar 2012.

\bibitem{PhysRevA.88.013605}
N.~K. Lowman and M.~A. Hoefer, ``Fermionic shock waves: Distinguishing
  dissipative versus dispersive regularizations,'' {\em Phys. Rev. A}, vol.~88,
  p.~013605, Jul 2013.

\bibitem{PhysRevLett.117.130402}
B.~Bertini and M.~Fagotti, ``Determination of the nonequilibrium steady state
  emerging from a defect,'' {\em Phys. Rev. Lett.}, vol.~117, p.~130402, Sep
  2016.

\bibitem{https://doi.org/10.48550/arxiv.2104.05812}
G.~A. El, ``Soliton gas in integrable dispersive hydrodynamics,'' 2021.

\bibitem{Petz1993-yb}
D.~Petz and G.~Toth, ``The bogoliubov inner product in quantum statistics,''
  {\em Lett. Math. Phys.}, vol.~27, pp.~205--216, Mar. 1993.

\bibitem{2206.14167}
B.~Doyon, G.~Perfetto, T.~Sasamoto, and T.~Yoshimura, ``Ballistic macroscopic
  fluctuation theory,'' 2020.
\newblock arXiv:2206.14167.

\bibitem{CortsCubero2019}
A.~C. Cubero and M.~Panfil, ``Thermodynamic bootstrap program for integrable
  {QFT}'s: form factors and correlation functions at finite energy density,''
  {\em Journal of High Energy Physics}, vol.~2019, Jan. 2019.

\bibitem{De_Nardis_2015}
J.~D. Nardis and M.~Panfil, ``Density form factors of the 1d bose gas for
  finite entropy states,'' {\em Journal of Statistical Mechanics: Theory and
  Experiment}, vol.~2015, p.~P02019, feb 2015.

\bibitem{10.21468/SciPostPhys.9.6.082}
E.~Granet and F.~H.~L. Essler, ``{A systematic $1/c$-expansion of form factor
  sums for dynamical correlations in the Lieb-Liniger model},'' {\em SciPost
  Phys.}, vol.~9, p.~82, 2020.

\bibitem{RevCorrelations}
J.~D. Nardis, B.~Doyon, M.~Medenjak, and M.~Panfil, ``Correlation functions and
  transport coefficients in generalised hydrodynamics,'' 2021.
\newblock arXiv:2104.04462.

\bibitem{Cubero2019}
A.~C. Cubero and M.~Panfil, ``{Generalized hydrodynamics regime from the
  thermodynamic bootstrap program},'' {\em SciPost Phys.}, vol.~8, p.~4, 2020.

\bibitem{PhysRev.40.749}
E.~Wigner, ``On the quantum correction for thermodynamic equilibrium,'' {\em
  Phys. Rev.}, vol.~40, pp.~749--759, Jun 1932.

\bibitem{Moyal1949}
J.~E. Moyal, ``Quantum mechanics as a statistical theory,'' {\em Mathematical
  Proceedings of the Cambridge Philosophical Society}, vol.~45, pp.~99--124,
  Jan. 1949.

\bibitem{PhysRevD.20.414}
P.~Sharan, ``Star-product representation of path integrals,'' {\em Phys. Rev.
  D}, vol.~20, pp.~414--418, Jul 1979.

\bibitem{Essler2022}
F.~H. Essler, ``A short introduction to generalized hydrodynamics,'' {\em
  Physica A: Statistical Mechanics and its Applications}, p.~127572, May 2022.

\bibitem{PhysRevLett.110.060602}
V.~Eisler and Z.~R\'acz, ``Full counting statistics in a propagating quantum
  front and random matrix spectra,'' {\em Phys. Rev. Lett.}, vol.~110,
  p.~060602, Feb 2013.

\bibitem{Viti2016}
J.~Viti, J.-M. St{\'{e}}phan, J.~Dubail, and M.~Haque, ``Inhomogeneous quenches
  in a free fermionic chain: Exact results,'' {\em {EPL} (Europhysics
  Letters)}, vol.~115, p.~40011, Aug. 2016.

\bibitem{Spohn2018}
H.~Spohn, ``Interacting and noninteracting integrable systems,'' {\em Journal
  of Mathematical Physics}, vol.~59, p.~091402, Sept. 2018.

\bibitem{PhysRevB.104.115423}
S.~Scopa, P.~Calabrese, and L.~Piroli, ``Real-time spin-charge separation in
  one-dimensional fermi gases from generalized hydrodynamics,'' {\em Phys. Rev.
  B}, vol.~104, p.~115423, Sep 2021.

\bibitem{PhysRevB.106.134314}
S.~Scopa, P.~Calabrese, and L.~Piroli, ``Generalized hydrodynamics of the
  repulsive spin-$\frac{1}{2}$ fermi gas,'' {\em Phys. Rev. B}, vol.~106,
  p.~134314, Oct 2022.

\bibitem{PhysRevLett.123.130602}
A.~Bastianello, V.~Alba, and J.-S. Caux, ``Generalized hydrodynamics with
  space-time inhomogeneous interactions,'' {\em Phys. Rev. Lett.}, vol.~123,
  p.~130602, Sep 2019.

\bibitem{DeNardis2019}
J.~{De Nardis}, D.~Bernard, and B.~Doyon, ``{Diffusion in generalized
  hydrodynamics and quasiparticle scattering},'' {\em SciPost Phys.}, vol.~6,
  p.~49, 2019.

\bibitem{Spohn1982-yq}
H.~Spohn, ``Hydrodynamical theory for equilibrium time correlation functions of
  hard rods,'' {\em Ann. Phys. (N. Y.)}, vol.~141, pp.~353--364, July 1982.

\bibitem{2004.07113}
T.~Yoshimura and H.~Spohn, ``Collision rate ansatz for quantum integrable
  systems,'' 2020.
\newblock arXiv:2004.07113.

\bibitem{CaoCradle}
X.~Cao, V.~B. Bulchandani, and J.~E. Moore, ``Incomplete thermalization from
  trap-induced integrability breaking: Lessons from classical hard rods,'' {\em
  Phys. Rev. Lett.}, vol.~120, p.~164101, Apr 2018.

\bibitem{Koch2022}
R.~Koch, J.-S. Caux, and A.~Bastianello, ``Generalized hydrodynamics of the
  attractive non-linear schrodinger equation,'' {\em Journal of Physics A:
  Mathematical and Theoretical}, vol.~55, p.~134001, Mar. 2022.

\bibitem{PhysRev.130.1605}
E.~H. Lieb and W.~Liniger, ``Exact analysis of an interacting bose gas. i. the
  general solution and the ground state,'' {\em Phys. Rev.}, vol.~130,
  pp.~1605--1616, May 1963.

\bibitem{1509.08332}
T.~Price and A.~Lamacraft, ``Quantum hydrodynamics in one dimension beyond the
  luttinger liquid,'' 2015.

\bibitem{You-reciprocal}
Z.~You, A.~Baskaran, and M.~C. Marchetti, ``Nonreciprocity as a generic route
  to traveling states,'' {\em PNAS}, vol.~117, pp.~19767--19772, 2020.

\bibitem{Fruchart-reciprocal}
M.~Fruchart, R.~Hanai, P.~B. Littlewood, and V.~Vitelli, ``Non-reciprocal phase
  transitions,'' {\em Nature}, vol.~592, no.~7854, pp.~363--369, 2021.

\bibitem{bratteli_operator_1987}
O.~Bratteli and D.~Robinson, {\em Operator {Algebras} and {Quantum}
  {Statistical} {Mechanics} 2: {Equilibrium} {States} {Models} in {Quantum}
  {Statistical} {Mechanics}}.
\newblock Operator {Algebras} and {Quantum} {Statistical} {Mechanics},
  Springer, 1997.

\end{thebibliography}
 
\end{document}